\documentclass[sigconf,preprint]{acmart}

\usepackage{booktabs}
\usepackage{dashrule}
\usepackage{array}
\usepackage{pifont}
\usepackage{multirow}
\usepackage{subcaption}
\usepackage{tikz}
\usepackage{xspace}
\usepackage{enumitem}
\usepackage{pgfplots}
\usepackage{tabularx}
\usepgfplotslibrary{colorbrewer}
\usepackage[most,breakable]{tcolorbox}
\tcbset{
  promptbox/.style={
    colback=blue!2!white,
    colframe=blue!60!black,
    boxrule=0.4pt, arc=2pt,
    left=8pt, right=8pt, top=6pt, bottom=6pt,
    fonttitle=\bfseries, breakable, enhanced
  }
}

\renewcommand\footnotetextcopyrightpermission[1]{}

\newcommand{\model}[0]{TCellAlign\xspace}

\begin{document}

\title{\model: Cross-study T-cell Populations Alignment with Nomenclature-Guided Multi-Agent Workflow}

\author{Pengyu Xie}
\affiliation{%
  \institution{The Chinese University of Hong Kong, Shenzhen}
  \department{School of Artificial Intelligence}
  \city{Shenzhen}
  \state{Guangdong}
  \country{China}}
\email{xiepengyu@cuhk.edu.cn}

\author{Rongjia Zhou}
\affiliation{%
  \institution{Emory University}
  \department{Department of Computer Science and Biology}
  \city{Atlanta}
  \state{GA}
  \country{USA}}
\email{rzhou73@emory.edu}

\author{Zhilin Ou}
\affiliation{%
  \institution{The Chinese University of Hong Kong, Shenzhen}
  \department{School of Artificial Intelligence}
  \city{Shenzhen}
  \state{Guangdong}
  \country{China}}
\email{ouuzhilin@gmail.com}

\author{Junyuan Zhang}
\affiliation{%
  \institution{The University of Melbourne}
  \department{Department of Biochemistry and Pharmacology}
  \city{Melbourne}
  \state{VIC}
  \country{Australia}}
\email{junyuan.zhang.1@student.unimelb.edu.au}

\author{Xiang Zhou}
\affiliation{%
  \institution{Yale University}
  \department{Department of Statistics and Data Science}
  \city{New Haven}
  \state{CT}
  \country{USA}}
\email{xiang.zhou.xz735@yale.edu}

\author{Xiaobo Sun}
\authornote{Corresponding authors: Xiaobo Sun, Jiaying Lu, and Wenjing Ma}
\affiliation{%
  \institution{Emory University}
  \department{Department of Human Genetics}
  \city{Atlanta}
  \state{GA}
  \country{USA}}
\email{xiaobo.sun@emory.edu}

\author{Jiaying Lu}
\authornotemark[1]
\affiliation{%
  \institution{Emory University}
  \department{Center for Data Science, School of Nursing}
  \city{Atlanta}
  \state{GA}
  \country{USA}}
\email{jiaying.lu@emory.edu}

\author{Wenjing Ma}
\authornotemark[1] 
\affiliation{%
  \institution{The Chinese University of Hong Kong, Shenzhen}
  \department{School of Artificial Intelligence}
  \city{Shenzhen}
  \state{Guangdong}
  \country{China}}
\email{wma2026@cuhk.edu.cn}

\renewcommand{\shortauthors}{Xie et al.}

\begin{abstract}
Cell type standardization plays a central role in integrating biological knowledge across single-cell studies. While standardized resources (\textit{e.g.,} Cell Ontology, Nomenclature Frameworks) provide unified vocabularies of cell populations, scientific publications and public datasets continue to use heterogeneous study-specific labels, making cross-study comparison difficult even when biologically equivalent cell populations are described. In this work, we are the first to formulate this challenge as an evidence-grounded cell population alignment problem and propose \model, a multi-agent framework that includes literature retrieval, information extraction, nomenclature-guided label alignment, and evidence-based adjudication. This modular design preserves the original terminology and supporting evidence reported by each study while producing standardized labels that can be compared across studies. We further construct a manually validated benchmark dataset linking study-specific labels, CZ CELLxGENE annotations, and standardized T-cell nomenclature across 44 manually curated, published studies (including over seven million cells) spanning four biological categories: healthy, cancer, infectious disease and inflammatory diseases. Across the evaluated tasks, \model achieves stronger semantic agreement than ontology-based baselines and maintains transcriptomic coherence with both open-source and closed-source large language models (LLM)  backbones. By connecting literature, datasets, and expert's nomenclature, \model enables consistent interpretation of T-cell subtypes and states across studies, facilitating biological knowledge integration and the development of future foundation models built upon standardized cellular representations.


\end{abstract}




\maketitle

\section{INTRODUCTION}

\begin{figure*}[t]
    \centering
    \includegraphics[width=0.9\textwidth]{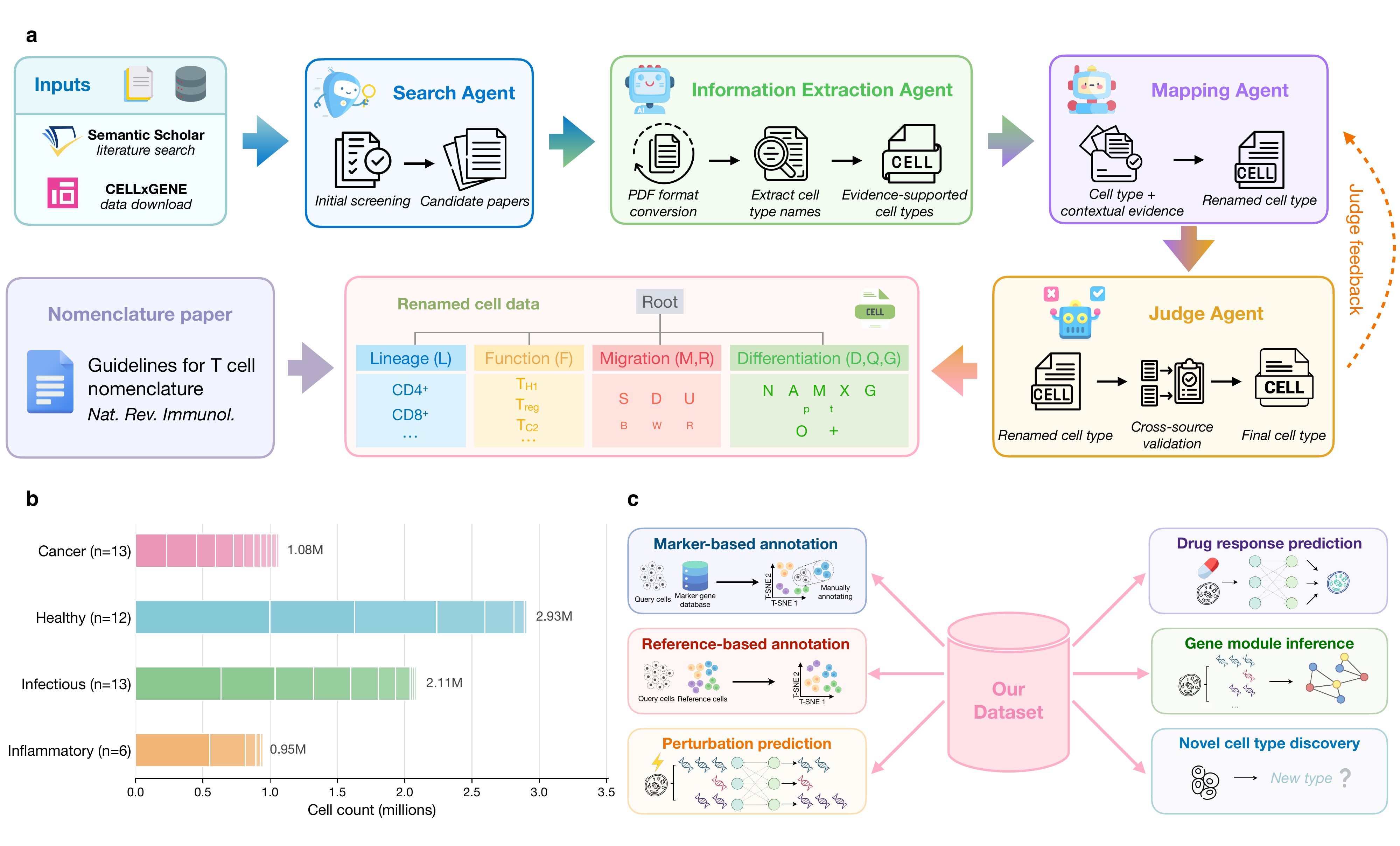}
    \vspace{-1em}
    \caption{Overview of our study. \textmd{(a) Overview of the proposed \model framework; (b) Statistics of our constructed benchmark dataset; (c) Potential biological applications for benchmark dataset after applying \model.}}
    \label{fig:workflow}
\vspace{-1em}
\end{figure*}

%
%
%
%



Cell types and cell states provide a fundamental basis for understanding biological development~\cite{cao2019organogenesis}, tissue homeostasis~\cite{lake2023kidneyatlas}, immune responses~\cite{wilk2020immuneatlas}, and disease mechanisms~\cite{chu2023pancancertcellatlas}. Recent advances in single-cell transcriptomics technologies have enabled increasingly fine-grained characterization of cellular subtypes and functional states, leading to a rapidly growing body of cell atlases across diverse biological systems~\cite{regev2017humancellatlas,han2018mousecellatlas,tabulamuris2018atlas}. However, as the resolution of cellular characterization increases, so does the diversity of terminology used to describe biologically related cell populations. Individual studies often introduce study-specific cell labels based on distinct experimental settings and functional interpretations, making it difficult to compare, integrate, and reuse across studies.


To facilitate consistent communication, the biomedical community has developed standardized ontologies~\cite{diehl2016cellontology, bard2005ontology, osumi2021cell}, manually-curated marker gene databases ~\cite{hu2023cellmarker, franzen2019panglaodb} and nomenclature systems~\cite{masopust2026tcellnomenclature, yuste2020community, spits2013innate} for unambiguously representing cell identities. Although promising in principle, real-world practice differs substantially. In practice, cell type identification is largely performed manually and does not utilize any above resources. A standard procedure is that after clustering cells based on their transcriptomic profiles, domain experts assign biological identities by examining canonical marker genes. Depending on the biological question, annotation may stop at broad cell categories or proceed to increasingly fine-grained subtypes through iterative rounds of subclustering. For example, Saunders \textit{et al.} ~\cite{saunders2018molecular} employed a two-stage clustering strategy that identified 565 neuronal populations in the adult mouse brain, assigning study-specific labels such as ``Frontal Cortex - Cluster 1 - SubCluster 5'' based on biological knowledge and independent component (IC) loadings derived by independent component analysis (ICA) from single-cell transcriptomics data. Consequently, critical discrepancies exist between standardized definitions and study-specific annotations, posing a major obstacle to large-scale cross-study integration and automated knowledge discovery.


Additionally, among the above resources, Cell Ontology (CL) stands out as a controlled vocabulary that describes cell types across tissues and species, promoting standardized and interoperable cell type annotations~\cite{tan2026cellontology}. Existing computational approaches primarily formulate cell type standardization  as an ontology matching problem, where lexical similarity or pretrained biomedical language models (PLMs) are used to map study-specific labels to standardized terms~\cite{robertson2009BM25, ristad1998editDist, zooma, bioportal, gu2022pubmedbert, liu2021SapBERT, wang2021onclass, ergen2024popv}. However, CL primarily provides a representation of canonical cell types and may not fully capture the increasingly fine-grained cellular states revealed by single-cell transcriptomics~\cite{wang2021onclass,tan2026cellontology}. This challenge is particularly evident for immune cells, where cellular populations are often defined not only by lineage but also by dynamic functional states. Among them, T-cell subsets represent a particularly challenging example, as their populations frequently encode multiple biological dimensions, including lineage, function, migration, differentiation and antigen status. Thus, community-driven T-cell nomenclature frameworks provide a complementary reference for harmonizing fine-grained T-cell populations reported in the literature ~\cite{masopust2026tcellnomenclature}. Moreover, standardized definitions available in curated single-cell data resources often provide limited context regarding the biological rationale, marker evidence, and nomenclature decisions underlying the assigned labels ~\cite{czcellxgene2025discover}. These contextual details are frequently preserved in the original scientific literature, suggesting that effective T-cell nomenclature alignment requires integrating structured annotations with evidence extracted from primary studies.


Motivated by recent efforts in establishing consensus T-cell nomenclature ~\cite{masopust2026tcellnomenclature}, which advocates a modular naming strategy for representing the complexity of T-cell biology, we formulate cross-study T-cell populations alignment as a multi-stage reasoning problem requiring literature understanding, biological evidence integration, and nomenclature-guided alignment. To address this challenge, we propose \model, a nomenclature-guided multi-agent framework for evidence-supported cross-study T-cell populations alignment (Figure ~\ref{fig:workflow}a). Rather than treating cell type standardization as a direct matching problem, \model decomposes the task into four specialized reasoning stages: study retrieval, evidence extraction, nomenclature mapping, and mapping adjudication. Specifically, a Search Agent identifies relevant studies and associated single-cell data datasets, an Information Extraction Agent extracts paper-specific T-cell labels together with supporting evidence, a Mapping Agent aligns labels to a consensus T-cell nomenclature, and a Judge Agent verifies ambiguous mappings by integrating nomenclature standards and biological evidence. 

Our work makes \textbf{three main contributions} from the perspectives of paradigm, resource, and methodology. \textbf{From a paradigm perspective}, we are the first to formulate cross-study T-cell nomenclature alignment as a knowledge integration problem that unifies standardized nomenclature, study-specific annotations, and biological evidence. \textbf{From a resource perspective}, we construct the first manually validated benchmark dataset that links paper-specific T-cell populations label, expert-curated cell ontologies ~\cite{czcellxgene2025discover}, and standardized T-cell nomenclature across 44 studies containing over seven million cells, providing a valuable community resource for cross-study knowledge integration and downstream applications such as reference-based annotation, novel T-cell state discovery, and gene regulatory network inference (Figure~\ref{fig:workflow}b,c). \textbf{From a methodological perspective}, we develop \model, a nomenclature-guided multi-agent framework that integrates literature retrieval, information extraction, nomenclature-guided mapping, and evidence-based adjudication to achieve robust and biologically interpretable cross-study T-cell alignment. Although demonstrated on T-cell nomenclature, the framework is readily extensible to other consensus nomenclature systems and literature-grounded biological knowledge integration tasks.

\section{RELATED WORK}

\subsection{Standards for T-Cell Population Alignment}

The need for consistent cell identity standardization is especially acute for T-cells, whose names often combine lineage, differentiation, activation, exhaustion, cytotoxicity, tissue residency, proliferation, clonality, and disease-context descriptors. As a result, biologically related T-cell populations are frequently described using heterogeneous terminology across studies. Several resources have been developed to facilitate standardized T-cell subtype alignment. \textbf{Cell Ontology} (CL)~\cite{bard2005ontology,diehl2016cellontology} provides a structured and hierarchical vocabulary for representing cell types across tissues and species. However, CL is primarily designed to organize cell identities through ontology relationships and may not fully capture the context-dependent functional states and compositional descriptors increasingly used in studies involving T-cell subtypes. For example, exhausted, tissue-resident, or progenitor-like T-cell subtypes are often defined by combinations of lineage, functional state, and environmental context rather than a single fixed cell type term. Complementary resources, such as \textbf{curated marker databases}~\cite{franzen2019panglaodb,hu2023cellmarker}, provide molecular evidence for supporting cell identity assignment but do not directly address inconsistencies in terminology across studies. To bridge these gaps, recent consensus efforts have proposed a standardized \textbf{T-cell nomenclature} based on compositional descriptors in a modular format, enabling a systematic representation of diverse T-cell states and functional phenotypes~\cite{masopust2026tcellnomenclature}. Compared with general-purpose ontologies, such nomenclature frameworks are better suited for harmonizing fine-grained T-cell subtypes reported in the literature.



\subsection{Computational Approaches to T-Cell Population Extraction and Alignment}

Beyond these standardization resources, computational approaches to cell identity can be grouped into three complementary lines of work. First, biomedical text-mining models such as BioBERT~\cite{lee2020biobert}, SciBERT~\cite{beltagy2019scibert}, and PubMedBERT~\cite{gu2022pubmedbert} support entity recognition, relation extraction, and other structured knowledge extraction tasks from scientific literature. However, these tasks typically treat entities or relations as their primary outputs, whereas fine-grained T-cell labels often include paper-specific compositional expressions whose interpretation depends on surrounding evidence. Second, cell-typing tools and resources such as CellTypist~\cite{dominguezconde2022celltypist}, ProjecTILs~\cite{andreatta2021projectils}, ImmCluster~\cite{jiang2023immcluster} and scPanKD ~\cite{ma2025scpankd} infer cell identities from expression profiles, marker genes, or reference atlases. Although effective for scalable annotation, their predictions are anchored to molecular inputs and predefined references, and do not directly reconcile heterogeneous names used for related T-cell states across papers. More recently, LLM-based approaches use biological knowledge to annotate cells or clusters from molecular evidence; scExtract also uses article-derived context, whereas CASSIA structures annotation through collaborating agents \cite{hou2024gpt4celltype,wu2025scextract,xie2025cassia}. These advances expand both evidence sources and reasoning mechanisms for cell typing, but their principal output remains a newly assigned cell identity rather than an evidence-linked mapping from an author-reported label to an explicit T-cell nomenclature. Thus, existing work separately advances literature extraction, expression-based typing, and biological reasoning, but does not yet integrate them into a unified process for traceable cross-study T-cell label alignment.

\subsection{Positioning of This Work}

Our work is positioned at the intersection of single-cell annotation, biomedical information extraction, and nomenclature-guided knowledge harmonization. Unlike conventional cell annotation methods that infer cell identities directly from expression profiles, \model focuses on the complementary but unresolved challenge of harmonizing heterogeneous labels reported across scientific studies. Unlike ontology resources that provide standardized vocabularies, \model does not assume that study-specific labels conform to predefined concepts; instead, it adopts a flexible framework proposed by the T-cell nomenclature while extracting and reasoning over the biological context underlying each label. Unlike general biomedical information extraction systems that focus on retrieving isolated entities or relations, \model addresses the more challenging task of aligning context-dependent T-cell population that combine lineage and functional states. Finally, unlike recent LLM-based single-cell annotation approaches that use language models as direct predictors of cell identities from molecular features, \model leverages LLMs as components of a structured literature understanding workflow, where specialized agents perform evidence extraction, nomenclature mapping, and adjudication.



\section{METHODS}


\subsection{Problem Definition}\label{sec:problem-definition}

Let $p\in\mathcal{P}$ denote a qualified scientific publication, where $\mathcal{P}$ is the set of source papers. Within each paper, let $O=\{o_1, o_2, \ldots, o_n\}$ denote the original label (i.e. cell type, cell state) named by its authors. For every original label $o_i$, let $\mathcal{E}_i$ denote the set of paper-local evidence spans supporting the biological meaning of that label. We define each ``label-evidence pair'' as $z_i=(o_i,\mathcal{E}_i)$. Given the T-cell nomenclature $\mathcal{N}$ proposed by Masopust et al.~\cite{masopust2026tcellnomenclature}, our goal is to map each $z_i$ to a nomenclature-guided standardized definition:
\begin{equation}
    f_{\mathcal N}(z_i)=y_i,
    \qquad
    y_i\in\mathcal{Y}_{\mathcal N},
\end{equation}
where $\mathcal{Y}_{\mathcal N}$ is the space of structured representations defined by $\mathcal{N}$. Specifically, $y_i$ is represented by seven atomic nomenclature slots indexed by $\mathcal K=\{L,F,M,R,D,Q,G\}$ corresponding to lineage, function, migration, migration subscript, differentiation state, differentiation state subscript, and antigen status, respectively. These slots are further organized into four biological dimensions:
\begin{equation}
    y_i=\left(L_i,F_i,(M_i,R_i),(D_i,Q_i,G_i)\right).
\end{equation}
Although organized into four biological dimensions, the seven atomic slots can be inferred and validated independently, enabling fine-grained representation of heterogeneous T-cell identities. 

Each atomic slot is represented as $y_{ik}=(c_{ik},v_{ik}), \quad k\in\mathcal K$, where $c_{ik}$ must be a nomenclature component defined by $\mathcal{N}$ and $v_{ik}$ denotes its corresponding biological meaning. When the original label and paper-local evidence do not support the corresponding property, the corresponding slot is left empty: $y_{ik}=(\varnothing,\varnothing)$. An empty slot indicates insufficient evidence and should be distinguished from an explicit nomenclature component representing an unknown value. For example, the migration module $\mathrm{U}$ explicitly denotes unknown migration, whereas an empty migration slot indicates that the available evidence does not support any migration statement. Additional contextual information including tissue source, disease context, species, molecular markers, and experimental conditions may be retained as supporting evidence, but are not encoded in $y_i$ unless $\mathcal{N}$ defines a corresponding module.

Finally, a deterministic rendering function converts the structured representations into the final aligned labels $M$:

\vspace{-1em}
\begin{equation}
    m_i=g(y_i;\mathcal{N}),
\end{equation}

where $g$ concatenates all non-empty nomenclature components according to the ordering rules defined by $\mathcal{N}$. Therefore, $f_{\mathcal{N}}$ determines the biological content of the aligned representation based on the original label and supporting evidence, whereas $g$ ensures a consistent nomenclature-compliant label format. The resulting alignment preserves traceability by maintaining the original label and evidence supporting each encoded biological property.

\subsection{\model Model Architecture}

As shown in Figure \ref{fig:workflow}, \model is a four-stage workflow for transforming heterogeneous T-cell labels reported in the scientific literature into evidence-grounded structured representations under a modular nomenclature. The workflow consists of a Search Agent, an Information Extraction (IE) Agent, a Mapping Agent, and a Judge Agent. The Search Agent identifies relevant publications with available single-cell transcriptomic data. The IE Agentextracts the author-reported T-cell labels and supporting evidence while preserving the original terminology. The Mapping Agent transforms each ``label-evidence pair'' into a nomenclature-guided structured representation. The Judge Agent validates the mapping consistency against both source evidence and nomenclature rules..

This staged decomposes cross-study T-cell subtype alignment into four interconnected challenges: literature discovery, evidence extraction, nomenclature-guided mapping, and evidence-based judge. Literature retrieval establishes the study corpus and associated data resources; information extraction links original labels with traceable publication evidence; mapping assigns evidence-supported biological attributes to predefined nomenclature slots; and judge validates each slot-level decision against both source evidence and nomenclature constraints. By preserving the original label, supporting evidence, and structured representation throughout the process, \model ensures that every aligned annotation remains traceable and biologically interpretable.

Instead of regenerating complete labels after validation, \model performs selective slot-level refinement. Multiple judges from Judge Agent evaluate each nomenclature slot, and only inconsistent mappings are returned to the Mapping Agent with feedback, while validated slots remain unchanged. When evidence is insufficient, the corresponding information is preserved as unresolved rather than inferred. After each refinement round, the updated representation is converted into an aligned label through the deterministic nomenclature function and re-evaluated until convergence or a predefined stopping criterion is reached (Algorithm~\ref{alg:def}).



\newcounter{tcellalgorithm}
\vspace{-0.5em}
\begin{table}[htbp!]
\centering
\refstepcounter{tcellalgorithm}
\label{alg:def}
\footnotesize
\begin{tabular}{@{}r@{\hspace{0.6em}}p{\dimexpr\columnwidth-2.6em\relax}@{}}
\toprule
\multicolumn{2}{@{}l}{\textbf{Algorithm \thetcellalgorithm}\quad \model} \\
\midrule
& \textbf{Input:} Literature source $\mathcal{S}$; eligibility criteria
$\Gamma$; nomenclature $\mathcal{N}$. \\
& \textbf{Control:} Maximum number of adjudication rounds $T_{\max}$. \\
& \textbf{Output:} Finalized records $\mathcal{J}$ containing structured
mappings, aligned labels, and adjudication histories. \\
\midrule
1 & $\mathcal{P}\leftarrow\operatorname{SearchAgent}(\mathcal{S},\Gamma)$ \\
2 & \textbf{for each} qualified paper $p\in\mathcal{P}$ \textbf{do} \\
3 & \hspace*{1.2em}$Z_p\leftarrow\operatorname{IEAgent}(p)$
\hfill $\triangleright$ Extract labels and supporting evidence \\
4 & \hspace*{1.2em}$Y_p^{(0)}\leftarrow
\operatorname{MappingAgent}(Z_p,\mathcal{N})$
\hfill $\triangleright$ Generate seven-slot representations \\
5 & \textbf{end for} \\

6 & $Z\leftarrow\{Z_p\}_{p\in\mathcal P},\quad
Y^{(0)}\leftarrow\{Y_p^{(0)}\}_{p\in\mathcal P}$ \\

7 & \textbf{for} $t\leftarrow0,\ldots,T_{\max}-1$
\textbf{ do} \\

8 & \hspace*{1.2em}$V^{(t)}
\leftarrow
\operatorname{JudgeAgent}
(Z,Y^{(t)},\mathcal{N})$
\hfill $\triangleright$ Evidence-based slot-level adjudication \\

9 & \hspace*{1.2em}\textbf{if}
$\operatorname{AllConfirmed}(V^{(t)})$
\textbf{ then break} \\

10 & \hspace*{1.2em}$C^{(t)}
\leftarrow
\{(i,k):
\operatorname{status}(V_{i,k}^{(t)})
=\texttt{rejected}\}$ \\

11 & \hspace*{1.2em}\textbf{if}
$C^{(t)}=\varnothing$
\textbf{ then break} \\

12 & \hspace*{1.2em}$Y^{(t+1)}
\leftarrow Y^{(t)}$
\hfill $\triangleright$ Preserve unflagged representations \\

13 & \hspace*{1.2em}$Y_{C}^{(t+1)}
\leftarrow
\operatorname{MappingAgent}
(Z_C,\mathcal{N},
Y_C^{(t)},V_C^{(t)})$
\hfill $\triangleright$ Revise rejected slots \\

14 & \hspace*{1.2em}\textbf{if}
$Y^{(t+1)}=Y^{(t)}$
\textbf{ then break} \\

15 & \textbf{end for} \\

16 & $M\leftarrow g(Y^{(t+1)};\mathcal{N})$
\hfill $\triangleright$ Generate aligned labels \\

17 & $\mathcal{J}\leftarrow
\operatorname{Finalize}
(\mathcal{P},Z,Y^{(0:t)},M,V^{(0:t)})$ \\

18 & \textbf{return} $\mathcal{J}$ \\
\bottomrule
\end{tabular}
\vspace{-1em}
\end{table}

\noindent \textbf{Search Agent.}
The Search Agent initiates the workflow from a literature source $\mathcal{S}$, which can be obtained from biomedical database queries or a user-provided collection of candidate publications. It retrieves accessible candidate papers and applies the study-level eligibility criteria $\Gamma$. A publication is retained only if it reports T-cell types, subtypes, states, or populations suitable for downstream extraction and its corresponding single-cell transcriptomic data can be verified as available through CZ CELLxGENE for downstream analysis. Eligible studies are then ranked by citation count to prioritize processing. The resulting qualified paper set $\mathcal{P}$ retains its metadata, source PDF, screening decision, and verified data reference $D_p$ for each study (Appendix~\ref{sec:supp-literature-data}).

\noindent \textbf{Information Extraction Agent.}
For each publication $p\in\mathcal{P}$, the Information Extraction (IE) Agent first converts the source PDF into a machine-readable full-text representation and then extracts author-reported T-cell labels together with the textual evidence required to interpret their biological meanings. The extraction unit is a biologically meaningful cell population rather than individual textual mentions or cluster indentifiers. Multiple mentions referring to the same population are consolidated and paper-specific abbreviations or cluster identifiers are resolved using definitions provided within the publication, while preserving the original terminology. Each extracted label $o_i$ is paired with a set of evidence spans $\mathcal{E}_i$, including marker gene descriptions, phenotypic descriptions, or other functional properties, to form $z_i=(o_i,\mathcal{E}_i)$. For each publication $p$, these pairs form $\mathcal{Z}_p=\{z_i\}^{n_p}_{i=1}$ and are provided to the Mapping Agent for nomenclature-guided alignment.

\noindent \textbf{Mapping Agent.}
Given a ``label-evidence pair'' $z_i=(o_i,\mathcal{E}_i)$, the Mapping Agent applies the nomenclature-guided mapping function $f_{\mathcal{N}}$ to infer an evidence-consistent structured representation: $y_i=f_{\mathcal{N}}(z_i)$. At refinement round $t$, the structured representation is denoted as $y_i^{(t)}=\{y_{ik}^{(t)}\}_{k\in\mathcal{K}}$, where $y_{ik}^{(t)}=(c_{ik}^{(t)},v_{ik}^{(t)})$ denotes the $k$-th atomic slot and $\mathcal{K}$ indexes the seven atomic nomenclature slots defined above. A slot is populated only when the original label $o_i$ or supporting evidence $\mathcal{E}_i)$ supports the corresponding biological attribute; otherwise, the slot remains empty.
After inferring all supported attributes, the final aligned label is generated deterministically with $m_i^{(t)}=g(y_i^{(t)};\mathcal{N})$. For publication $p$, we denote $Y_p^{(t)}=\{y_i^{(t)}\}_{i=1}^{n_p}$ and $M_p^{(t)}=\{m_i^{(t)}\}_{i=1}^{n_p}$. This separation between structured inference and label generation ensures that each mapped label is generated from an explicit and interpretable biological representation. Each record retains the original label $o_i$, supporting evidence $\mathcal{E}_i)$, structured representation $y_i$, mapped label $m_i$, and nomenclature $\mathcal{N}$. During later refinement rounds, only slots identified by the Judge Agent as inconsistent are updated, while validated slots remain unchanged.

\noindent \textbf{Judge Agent and Iterative Refinement.}
At refinement round $t$, the Judge Agent independently assesses whether every slot in $y_i^{(t)}$ is consistent with the original label $o_i$, paper evidence $\mathcal{E}_i$, mapped label $m_i^{(t)}$, and nomenclature constraints $\mathcal{N}$. Three independent judges evaluate each slot, and the majority decision is recorded as either \texttt{confirmed} or \texttt{rejected}. The resulting records are denoted by $V^{(t)}$ with $V_{p,i,k}^{(t)}$ representing the validation outcome for slot $k$ of label $i$ in publication $p$.
During refinement, rejected slots requiring revision are returned to the Mapping Agent with Judge feedback, while confirmed slots are preserved. The Mapping Agent then updates the flagged slots based on the original ``label-evidence'' pair, nomenclature constraints, and feedback, after which $m_i^{(t)}$ is regenerated and re-evaluated. Claims that cannot be verified from the available evidence remain unresolved rather than being filled. The iterative refinement process terminates when all mappings are confirmed, no actionable feedback remains, the mapping converges, or the maximum number of refinement rounds $T_{\max}$ is reached. All intermediate mappings and adjudication records are retained to make the refinement process traceable and reviewable.

\subsection{Benchmark Dataset Construction}\label{sec:benchmark-construction}

To evaluate cross-study T-cell populations alignment, we constructed a manually validated benchmark database linking the paper-reported T-cell subsets with both author-provided annotations and the standardized Cell Ontology (CL) annotations available in the released single-cell datasets ~\cite{czcellxgene2025discover}. Following the data collection procedure described in Appendix~\ref{sec:supp-literature-data}, we screened candidate publications with Search Agent and retained 44 studies covering four biological categories: Healthy, Cancer, Infectious and Inflammatory (Appendix Table~\ref{tab:supp-per-study-summary}). Each retained study  contains a T-cell-relevant H5AD dataset, an author-provided annotation list, and a manually verified correspondence between paper-reported labels and data annotations (Appendix~\ref{sec:supp-paper-data-matching}). These 44 studies constituted the benchmark used for the subsequent alignment evaluation.

For each retained study, our IE Agent first extracts the set of paper-provided T-cell labels $O=\{o_1, o_2, \ldots, o_n\}$. With manual curation (Appendix~\ref{sec:supp-paper-data-matching}), we obtained $O\rightarrow A$, where $A=\{a_1, a_2, \ldots, a_m\}$ denotes the set of author-provided annotations. This manual curation ensures that the benchmark explicitly connects terminology used in the original publications with the annotations provided in the released datasets. In addition, CZ CELLxGENE provides standardized CL annotations through the \texttt{cell\_type} field and associated ontology term identifiers \texttt{cell\_type\_ontology\_term\_id}. We associate each data annotation with its corresponding CL term(s), yielding $A\rightarrow CL$ and the resulting correspondence between study-specific terminology and standardized cell identity representations, denoted by $O \rightarrow A \rightarrow CL$.

Given extracted paper labels $O$, each alignment method, including TCellAlign and competing methods such as PubMedBERT ~\cite{gu2022pubmedbert}, ZOOMA ~\cite{zooma}, BioPortal ~\cite{bioportal}, etc., predicts a mapped label $O \rightarrow M$, where $M=\{m_1, m_2, \ldots, m_n\}$ denotes the mapped labels generated by the method. Because the original paper labels $O$ serve as a shared intermediate representation, we establish an induced correspondence between predicted labels $M$ and reference CL annotations with $M \leftarrow O \rightarrow A \rightarrow CL$. This correspondence enables quantitative evaluation of alignment quality from two complementary perspectives: (i) annotation-level consistency by comparing predicted label groupings with CL-based reference groupings, and (ii) biological coherence by evaluating whether cells assigned to the same predicted label exhibit consistent transcriptomic profiles. Additional details regarding the baseline methods and evaluation metrics are provided in Appendices~\ref{apex:baselines} and~\ref{apex:evaluation-metrics}.

\begin{table*}[htbp!]
\centering
\small
\caption{Performance comparison across four datasets. Higher is better for all metrics. AMI: Adjusted Mutual Information. F1: Merge F1. MC: Merge Consistency. \textmd{\textit{Note:} Within each IE block, \textbf{bold} and \underline{underlined} values mark the column-best and second-best results, respectively. Underlining is omitted for tied maxima. $^\dagger$ denotes thinking mode.}}
\label{tab:exp-results}
\vspace{-1em}
\setlength{\tabcolsep}{3pt}
\begin{tabular}{@{}cl|ccc|ccc|ccc|ccc|ccc@{}}
\toprule
\multicolumn{2}{c|}{} & \multicolumn{3}{c|}{\textbf{All}}
& \multicolumn{3}{c|}{Healthy}
& \multicolumn{3}{c|}{Cancer}
& \multicolumn{3}{c|}{Infectious}
& \multicolumn{3}{c}{Inflammatory} \\
\cmidrule(lr){3-5}
\cmidrule(lr){6-8}
\cmidrule(lr){9-11}
\cmidrule(lr){12-14}
\cmidrule(lr){15-17}

IE & Method
& AMI &  F1 & MC
& AMI &  F1 & MC
& AMI &  F1 & MC
& AMI &  F1 & MC
& AMI &  F1 & MC\\
\midrule

\multirow{7}{*}{\textit{GPT-5.5}}
& BM25 & 0.254 & 0.347 & 0.264 & 0.427 & 0.511 & 0.313 &  0.206 & 0.184 & \underline{0.226} &  0.201 & 0.122 & 0.191 & 0.120 & 0.296 & 0.367 \\

& Levenshtein & 0.368 & 0.350 & 0.352 & 0.311 & 0.463 & 0.381 & 0.381 & 0.211 & 0.215 & 0.447 & NA & \textbf{0.572} & 0.286 & 0.338 & 0.226  \\

& PubMedBERT & 0.518 & 0.479 & \textbf{0.407} &  0.424 & 0.525 & 0.446 & 0.441 & 0.116 & 0.194 & 0.628 & 0.118 & 0.226 & \textbf{0.623} & \underline{0.705} & \textbf{0.753}  \\

& SapBERT & 0.524 & 0.439 & 0.345 & 0.447 & 0.590 & 0.399 & 0.417 & 0.022 & 0.060 & \underline{0.767} & NA & 0.261 & 0.368 & 0.376 & \underline{0.641} \\

& BioPortal & \underline{0.532} & \underline{0.519} & 0.361 &  \textbf{0.775} & \underline{0.734} & \textbf{0.537} & \textbf{0.535} & 0.291 & 0.146 & 0.403 & 0.133 & \underline{0.514} & 0.321 & 0.514 & 0.265 \\

& ZOOMA & 0.197 & 0.381 & 0.252 & 0.285 & 0.422 & 0.295 & 0.280 & \textbf{0.381}  & 0.161 & -0.023 & \textbf{0.357} & 0.324 & 0.330 & 0.339 & 0.218\\

& \textbf{\model} & \textbf{0.647} & \textbf{0.586} & \underline{0.378} &
\underline{0.671} & \textbf{0.845} & \underline{0.460} & \underline{0.474} & \underline{0.304} & \textbf{0.320} & \textbf{0.826} & \underline{0.337} & 0.434 & \underline{0.558} & \textbf{0.800} & 0.345 \\
\midrule

\multirow{7}{*}{\shortstack{\textit{DeepSeek-}\\\textit{V4-Pro}$^\dagger$}}
& BM25 & 0.251 & 0.428 & 0.273 & 0.347 & 0.562 & 0.253 & 0.302 & 0.350 & 0.285 & 0.165 & 0.377 & 0.304 & 0.146 & 0.369 & 0.235 \\

& Levenshtein & 0.340 & 0.469 & 0.286 & 0.395 & \underline{0.707} & 0.336 & 0.283 & \textbf{0.394} & 0.306 & 0.362 & 0.362 &  0.146 & 0.295 & 0.322 & 0.520\\

& PubMedBERT &  \underline{0.620} & \underline{0.579} & \textbf{0.511} & 0.532 & 0.703 & \underline{0.453} & \textbf{0.476}& 0.376 &  \textbf{0.441} & \textbf{0.847}& 0.528 & 0.253 & \underline{0.591} & 0.643 & \underline{0.749} \\

& SapBERT &  0.544& 0.374  & \underline{0.460} & 0.527& 0.643 & \textbf{0.472} & 0.382& 0.277 & 0.436 & 0.783& 0.110 & 0.174 & 0.384 & 0.435 & \textbf{0.761}\\

& BioPortal & 0.504 & \textbf{0.589} & 0.396 & \underline{0.643} & 0.685 & 0.400 & \underline{0.433}& \underline{0.389} & 0.335 & 0.375& \textbf{0.570}  & \textbf{0.517} & \textbf{0.646} & \underline{0.753} & 0.223 \\

& ZOOMA &  0.241&  0.374 & 0.243 & 0.461 & 0.470  & 0.343 & 0.152 & 0.293 & 0.070 & 0.144& 0.380  & 0.285 & 0.188 & 0.366 & 0.327\\

& \textbf{\model} &  \textbf{0.636}&  0.535  & 0.426 & \textbf{0.685} & \textbf{0.814}& 0.354 & 0.428 & 0.209 & \textbf{0.441} & \underline{0.822}& \textbf{0.570} & \underline{0.367} & 0.553& \textbf{0.778} & 0.616\\
\bottomrule
\end{tabular}
\vspace{-1em}
\end{table*}

\section{RESULTS}
Overall, we investigate the following research questions. \textbf{RQ1:} Can \model generate label groupings that are consistent with the Cell Ontology annotations provided by CZ CELLxGENE? \textbf{RQ2:} Do the aligned labels correspond to biologically meaningful transcriptional populations compared to existing methods? \textbf{RQ3:} How does \model support biological interpretation? \textbf{RQ4:} How do individual technical components influence model performance?

\subsection{Experimental Settings}
\noindent \textbf{Compared baselines.} We adopt the following established cross-study cell type alignment methods as baselines for comparison. (1) Lexical similarity-based methods that map original cell names to Cell Ontology (CL) terms via string matching: BM25~\cite{robertson2009BM25} and Levenshtein distance~\cite{ristad1998editDist}. (2) Semantic similarity-based methods that leverage pre-trained language model embeddings: PubMedBERT~\cite{gu2022pubmedbert} and SapBERT~\cite{liu2021SapBERT}. (3) Bioinformatics tools that perform multi-ontology-based annotation: BioPortal~\cite{bioportal} and ZOOMA~\cite{zooma}. Further technical details on these baselines are given in Appendix~\ref{apex:baselines}.

\noindent \textbf{\model implementation details.} To evaluate the effect of different LLM backbones (RQ4), we first conduct component-wise ablation studies using six representative models, including the open-source models MiMo-V2.5-Pro, DeepSeek-V4-Pro, and DeepSeek-V4-Flash~\cite{xiaomi2026mimov25,xu2026deepseekv4}, and the closed-source models Gemini 3.5 Flash, Claude Opus 4.8, and GPT-5.5~\cite{deepmind2026gemini35flash,anthropic_2026_claude_opus_48,openai_2026_gpt55}. We additionally compare thinking and non-thinking variants of DeepSeek for Information Extraction, and of both DeepSeek and MiMo for Mapping. For Judge Agent ablations, the Mapping Agent is instantiated with GPT-5.5, DeepSeek-V4-Pro, or DeepSeek-V4-Flash, while the Judge ensemble is fixed to DeepSeek-V4-Pro, GLM-5.2~\cite{zeng2026glm5}, and GPT-5.5. Detailed configurations are presented in the corresponding ablation sections. Based on the ablation results, we instantiate the complete \model workflow using GPT-5.5 and DeepSeek-V4-Pro (thinking), the best-performing closed-source and open-source backbones, respectively. \model aligns T-cell populations to a standardized T-cell nomenclature through evidence-grounded information extraction, nomenclature-guided mapping, and evidence-based adjudication. The main experiments evaluate 44 studies comprising over seven million cells and both implementations use identical paper inputs, prompt templates, nomenclature rules, and adjudication protocols, differing only in the underlying LLM backbone.

\begin{figure*}[htbp!]
    \centering
    \includegraphics[width=0.9\textwidth]{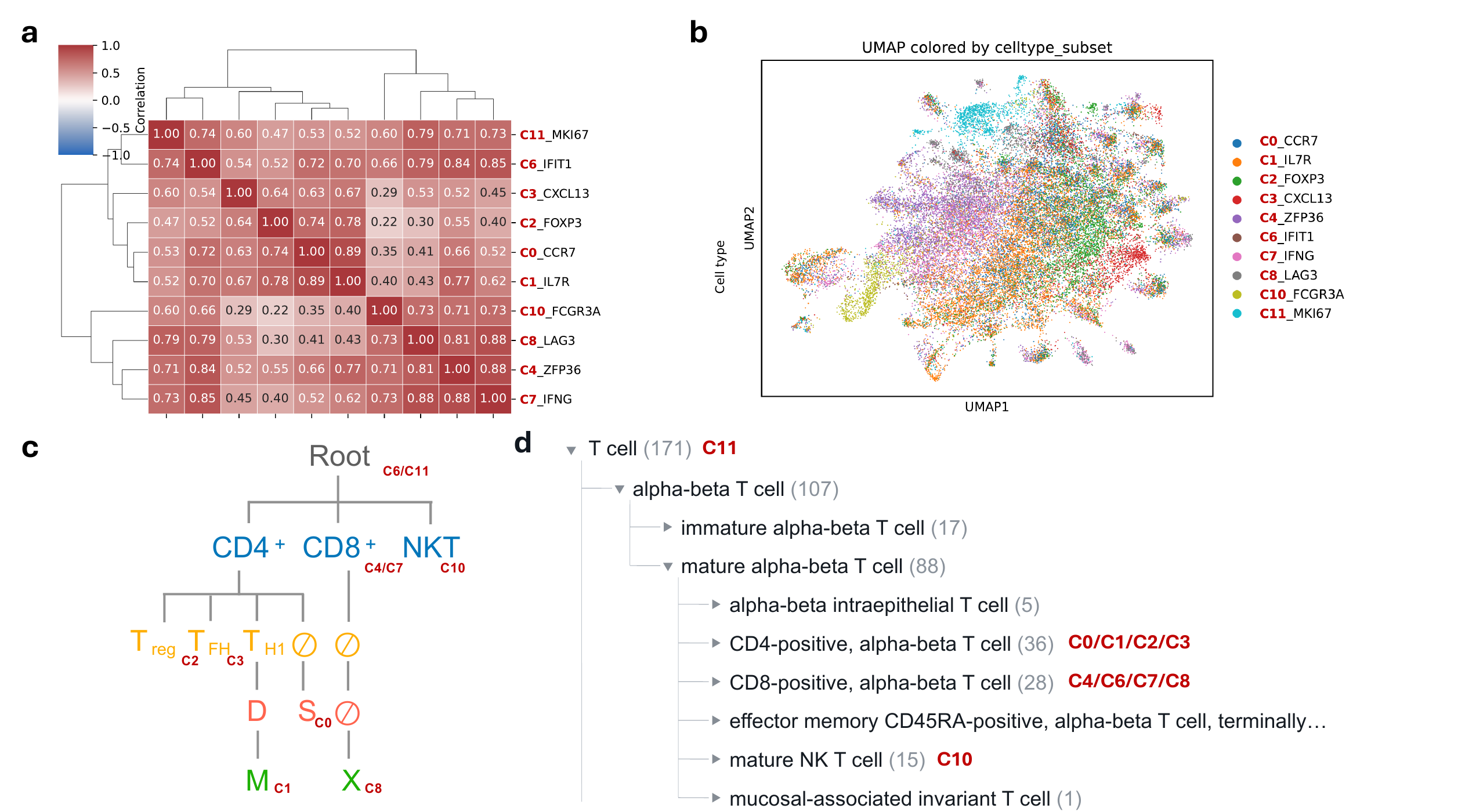}
    \vspace{-1em}
    \caption{Breast Cancer Case Study with Cluster Identifiers annotated. \textmd{(a) Clustered Heatmap. (b) UMAP visualization. (c) \model nomenclature-guided standardized definitions. (d) Cell Ontology standardized definitions.}}
    \label{fig:similarity}
\vspace{-1em}
\end{figure*}

\subsection{Grouping consistency evaluation (RQ1)}

As described in Section~\ref{sec:benchmark-construction}, we evaluate whether the label merging decisions produced by different methods are consistent with the Cell Ontology annotations provided by CZ CELLxGENE~\cite{czcellxgene2025discover}. Given the correspondence $M \leftarrow O \rightarrow A \rightarrow CL$, the reference CL labels ($CL$) and aligned labels ($M$) are merged partitions of the original data annotations ($A$). We therefore evaluate consistency between the predicted and reference grouping structures using Adjusted Mutual Information (AMI) and Merge F1 (Appendix~\ref{apex:evaluation-metrics}). As shown in Table~\ref{tab:exp-results}, \model with GPT-5.5 achieves the highest overall AMI and Merge F1 among all compared methods. Replacing GPT-5.5 with DeepSeek-V4-Pro (thinking) yields highly comparable results, with overall AMI and Merge F1 decreasing by only 0.011 and 0.051, respectively. The two implementations exhibit nearly identical AMI across all  categories, indicating that they recover largely consistent grouping structures. The largest difference is observed in the Cancer cohort, where DS-V4-Pro yields a lower Merge F1. We attribute this reduction primarily to its tendency to extract fewer paper-specific T-cell labels during the IE stage, producing coarser predicted partitions and consequently lower merge-level resolution. Overall, these results demonstrate that \model remains robust to the choice of LLM backbone.

\subsection{Transcriptomic coherence evaluation (RQ2)}

Using gene expression profiles from the CZ CELLxGENE datasets, we further evaluated whether each alignment method produces transcriptionally coherent cell populations. As shown in Table~\ref{tab:exp-results}, \model instantiated with GPT-5.5 achieves the second-highest Merge Consistency (MC), with PubMedBERT ranking first. We hypothesize that this difference reflects the two approaches' distinct merging strategies. Semantic similarity-based methods such as PubMedBERT and SapBERT primarily rely on local lexical or semantic similarity, and therefore tend to assign paper-specific labels to distinct ontology terms rather than merging them. Consequently, they perform fewer merges (0.727 and 0.818 per study on average, respectively), reducing the likelihood of combining transcriptionally heterogeneous populations and resulting in higher transcriptomic coherence. In contrast, \model performs evidence-guided alignment by jointly considering paper-derived biological evidence and the modular T-cell nomenclature. This produces a higher average merge rate (1.091 merges per study), yielding substantially better agreement with Cell Ontology (RQ1) while occasionally merging biologically related but transcriptionally distinguishable populations. The complementary behavior observed in RQ1 and RQ2 therefore suggests an inherent trade-off between ontology consistency and transcriptomic coherence. To better understand this trade-off, we further investigate representative alignment cases in the following section.




\subsection{Structural consistency evaluation (RQ3)}

Besides the quantitative evaluations, we performed a qualitative case study to examine whether the labels produced by \model preserve biologically meaningful structural relationships among T-cell populations. As discussed in Section~\ref{sec:problem-definition}, our nomenclature represents T-cell standardized definitions with four biological dimensions covering \textit{Lineage}, \textit{Function}, \textit{Migration}, and \textit{Differentiation and Antigen Status}. An ideal alignment to nomenclature should therefore not only assign consistent names, but also group biologically related cell populations under the same dimension while preserving their functional distinctions.

We selected a representative breast cancer study~\cite{wu2021breastatlas}, which provides diverse T-cell populations with detailed descriptions in the original publication (Section \textit{Lymphocytes and Innate Lymphoid Cells}) and corresponding validated dataset annotations. Using a GPT-5.5 backbone, the IE Agent extracted 10 T-cell populations corresponding to 12 annotated clusters (c0-c11) in the released dataset. Two clusters were excluded: cluster 5 (\texttt{c5\_CD8+\_GZMK}) was not explicitly described in the manuscript, and cluster 9 (\texttt{c9\_NK\_cells\_\\AREG}) represented natural killer (NK) cells rather than natural killer T (NKT) cells. Although NK and NKT cells share cytotoxic characteristics, they represent distinct immune populations, with only NKT cells included in our T-cell scope~\cite{liu2021nknkt}. After validation, we established correspondence between paper-reported labels and data annotations using cluster identifiers. Combining this correspondence with CZ CELLxGENE CL annotations enabled direct comparison between \model-generated labels and standardized ontology terms. The resulting mappings are summarized in Table~\ref{tab:case-study}.

Overall, the labels generated by \model exhibited structural organization consistent with both Cell Ontology annotations and transcriptomic similarity patterns measured by hierarchical clustering and UMAP visualization (Figure~\ref{fig:similarity}(a), (b)). For example, Clusters 0-3 were consistently assigned to the CD4$^{+}$ lineage while preserving distinct functional states, including T\textsubscript{reg} and T\textsubscript{FH}, and displayed highly similar transcriptional profiles. Similarly, Clusters 4, 7, and 8 were mapped to the CD8$^{+}$ lineage (Figure~\ref{fig:similarity}(c)) and formed a transcriptionally coherent group (Figure~\ref{fig:similarity}(a)). These results suggest that the proposed nomenclature alignment preserves both lineage relationships and functional distinctions among T-cell populations.

An informative exception was Cluster 6 (\texttt{c6\_IFIT1}). While CZ CELLxGENE annotated this cluster as \texttt{CD8-positive, alpha-beta T cell} (Figure~\ref{fig:similarity}(d)), the original study reported that the IFN-I signature cluster (\texttt{T cells:IFIT1/c6}) and the proliferative cluster (\texttt{T cells:MKI67/c11}) contained mixed CD4$^{+}$ and CD8$^{+}$ T-cell populations. Based on this evidence, the Mapping Agent assigned both clusters to the broader \textit{T-cell} lineage rather than imposing a CD4$^{+}$ or CD8$^{+}$ designation unsupported by the source literature. These two clusters also exhibited similar transcriptional profiles, consistent with the biological interpretation from the manuscript. This example highlights that \model does not merely reproduce existing ontology annotations, but integrates publication-specific evidence to produce biologically grounded nomenclature alignment.

\subsection{Ablation Study (RQ4)}

To examine three component-level design choices in \model, we evaluate paper-specific label extraction, alternative reference inputs for mapping, and whether refinement reduces mapping errors.


\begin{table*}[htbp!]
  \centering
  \caption{Information Extraction Agent performance against expert annotations. \textmd{\textit{Note:} Model names are abbreviated; full configurations are described in Experimental Settings. DS denotes DeepSeek. The Human column reports mean pairwise SBMS-F1 over six expert pairs per paper, macro-averaged across seven papers. \textbf{Bold} numeric values denote the highest non-human score in each row. $^\dagger$ denotes thinking mode.}}
  \vspace{-1em}
  \label{tab:stage1-prompt-v2}
  \begin{tabular*}{\textwidth}{@{\extracolsep{\fill}}lccccccccccc@{}}
    \toprule
    \multirow{2}{*}{SBMS-F1} & \multicolumn{5}{c}{Open-source models} &
    \multicolumn{3}{c}{Closed-source models} & \multicolumn{3}{c}{Baselines / reference} \\
    \cmidrule(lr){2-6}\cmidrule(lr){7-9}\cmidrule(lr){10-12}
    & MiMo-Pro & DS-Pro$^\dagger$ & DS-Flash$^\dagger$ & DS-Pro & DS-Flash &
    Gemini & Claude & GPT & TCell-6 & TCell-43 & Human \\
    \midrule
    Label only & \textbf{0.8954} & 0.8757 & 0.8620 & 0.8430 & 0.8475 &
    0.8599 & 0.8343 & 0.8407 & 0.5818 & 0.3539 & 0.9197 \\
    Label+Evidence & 0.8443 & 0.8373 & \textbf{0.8484} & 0.8315 & 0.8366 &
    0.8449 & 0.8287 & 0.8209 & 0.4373 & 0.2721 & 0.8838 \\
    \bottomrule
  \end{tabular*}
\vspace{-1em}
\end{table*}

\noindent \textbf{Information Extraction Agent.}
To evaluate paper-specific label and evidence extraction, we compare the IE Agent's outputs with annotations from four independent domain experts on the seven-paper evaluation set listed in Appendix Table~\ref{tab:supp-stage1-paper-inventory}. Because system outputs and expert annotations are variable-length sets of free-text terms, we use the \textit{Soft Bipartite Matching F1 Score} (\textbf{SBMS-F1}) for label-only and label+evidence evaluation. We compare the LLM settings with two fixed-nomenclature baselines and report human-human agreement as a descriptive reference (Appendix~\ref{appex:SBMS-F1}). All paper-contextualized LLM settings outperform the two fixed-nomenclature baselines on this evaluation set (Table~\ref{tab:stage1-prompt-v2}). MiMo-V2.5-Pro performs best for label-only extraction, whereas DeepSeek-V4-Flash (thinking) performs best for label-plus-evidence extraction, closely followed by Gemini 3.5 Flash. Moreover, the thinking variants score higher on both targets for the two DeepSeek models.

\vspace{-1.5em}
\begin{table}[htbp!]
  \centering
  \caption{Mapping Agent performance and efficiency with PDF and structured-rule references. \textmd{\textit{Note:} Gold-standard information is withheld from both reference inputs. Accuracy is shown as PDF$\rightarrow$Rule; $\Delta$ is Rule minus PDF, and savings are relative to PDF. $^\dagger$ denotes thinking mode.}}
  \vspace{-1em}
  \label{tab:stage2-normalization-ablation}
  \setlength{\tabcolsep}{2.5pt}
  \begin{tabular*}{\columnwidth}{@{\extracolsep{\fill}}lcccc@{}}
    \toprule
    \multirow{2}{*}{Model} & Acc. (\%) & \multirow{2}{*}{$\Delta$ (pp)} &
    \multicolumn{2}{c}{Saving (\%)} \\
    \cmidrule(lr){2-2}\cmidrule(lr){4-5}
    & PDF$\rightarrow$Rule & & Token & Cost \\
    \midrule
    \multicolumn{5}{l}{\textit{(Open-source models)}} \\
    \addlinespace[2pt]
    MiMo-V2.5-Pro$^\dagger$ & 81.2$\rightarrow$81.2 & $0.0$ & 66.3 & 54.6 \\
    MiMo-V2.5-Pro & 93.8$\rightarrow$81.2 & $-12.5$ & 76.2 & 69.2 \\
    DeepSeek-V4-Flash$^\dagger$ & 81.2$\rightarrow$81.2 & $0.0$ & 51.3 & 34.7 \\
    DeepSeek-V4-Flash & 62.5$\rightarrow$81.2 & $+18.8$ & 75.3 & 67.5 \\
    DeepSeek-V4-Pro$^\dagger$ & 93.8$\rightarrow$87.5 & $-6.3$ & 54.4 & 37.9 \\
    DeepSeek-V4-Pro & 75.0$\rightarrow$81.2 & $+6.3$ & 75.3 & 67.6 \\
    \midrule
    \multicolumn{5}{l}{\textit{(Closed-source models)}} \\
    \addlinespace[2pt]
    GPT-5.5 & 93.8$\rightarrow$93.8 & $0.0$ & 56.9 & 30.1 \\
    Claude Opus 4.8 & 87.5$\rightarrow$87.5 & $0.0$ & 72.3 & 45.4 \\
    Gemini 3.5 Flash & 93.8$\rightarrow$87.5 & $-6.3$ & 61.0 & 28.2 \\
    \bottomrule
  \end{tabular*}
\vspace{-1em}
\end{table}

\noindent \textbf{Mapping Agent.}
We assess alternative reference inputs for the Mapping Agent through a paired comparison of the paper PDF and a structured normalization rule on the 16-case gold-standard set in Table~7 of Masopust et al.~\cite{masopust2026tcellnomenclature}. To prevent answer leakage, we remove Table~7 from the PDF and the corresponding gold-standard examples from the rule, using the withheld mappings only for evaluation. For each model setting, we evaluate case-insensitive exact-match accuracy and record token count and estimated cost using public OpenRouter list prices (Table~\ref{tab:stage2-normalization-ablation}). The PDF and structured rule achieve nearly identical average exact-match accuracy across model settings, while the rule substantially reduces token use and estimated cost in every setting. We therefore use the structured rule in the subsequent Judge experiment as the more efficient reference input.

\noindent \textbf{Judge Agent.}
To evaluate the corrective effect of the Judge Agent, we compare the initial mapping accuracy and final accuracy after up to three rounds of Judge-guided refinement on the same 16-case ground truth dataset. All settings use the same structured rule without examples, evaluation set, and fixed Judge ensemble with only the Mapping Agent backend varies. Judge-guided refinement corrects one additional case for each backend, corresponding to a 6.25-percentage-point improvement: GPT-5.5 improves from 93.75\% to 100.00\%, while DeepSeek-V4-Pro and DeepSeek-V4-Flash in thinking mode improve from 87.50\% to 93.75\% and from 81.25\% to 87.50\%, respectively. All settings converge after the first refinement round. These results demonstrate that Judge-guided refinement consistently improves mapping accuracy across different Mapping Agent backends, while the final performance remains dependent on the quality of the initial mapping model.

\section{CONCLUSION}

We present \model, a nomenclature-guided multi-agent framework that integrates literature retrieval, information extraction, nomenclature-guided mapping, and evidence-based adjudication to align heterogeneous T-cell populations across studies. Experiments on 44 studies demonstrate that \model achieves superior evaluation performance, supporting transparent and biologically grounded cross-study T-cell nomenclature alignment. In the future, we plan to further improve the reasoning capabilities of the framework and expand the data resource to enable immunological discovery.


\section{Limitations and Ethical Considerations}
\noindent \textbf{Limitations.} While \model demonstrates strong performance in cross-study T-cell nomenclature alignment, several limitations remain. First, the benchmark includes only studies with accessible publications from CZ CELLXGENE,  which may limit coverage and introduce selection bias. Second, CL annotations as evaluation may not capture fine-grained or context-dependent T-cell populations. Lastly, our ablation studies are limited and broader validation across additional studies and nomenclature frameworks is needed to establish generalizability.

\noindent \textbf{Ethical Considerations.} Our study relies exclusively on publicly available, anonymized biomedical literature and single-cell transcriptomics datasets. Because no personally identifiable information is present or collected, the work raises no privacy concerns. The resulting computational resource for cell-type nomenclature mapping is intended solely for basic scientific research and should not be applied in clinical decision-making.

\section{Generative AI Usage}
Generative AI models were used to develop and evaluate \model, as described in the Methods and Experimental Settings, and for manuscript language editing. The authors reviewed and verified all AI-generated outputs and take full responsibility for the content of this manuscript.


\clearpage
\bibliographystyle{ACM-Reference-Format}
\bibliography{refs}

@article{dominguezconde2022celltypist,
  title = {Cross-tissue immune cell analysis reveals tissue-specific features in humans},
  author = {Dominguez Conde, Cecilia and others},
  journal = {Science},
  volume = {376},
  number = {6594},
  pages = {eabl5197},
  year = {2022},
  doi = {10.1126/science.abl5197}
}

@article{andreatta2021projectils,
  title = {Interpretation of T cell states from single-cell transcriptomics data using reference atlases},
  author = {Andreatta, Massimo and Corria-Osorio, Jesus and Muller, Soren and Cubas, Rafael and Coukos, George and Carmona, Santiago J.},
  journal = {Nature Communications},
  volume = {12},
  pages = {2965},
  year = {2021},
  doi = {10.1038/s41467-021-23324-4}
}

@article{jiang2023immcluster,
  title = {{ImmCluster}: an ensemble resource for immunology cell type clustering and annotations in normal and cancerous tissues},
  author = {Jiang, Tiantongfei and Zhou, Weiwei and Sheng, Qi and Yu, Jiaxin and Xie, Yunjin and Ding, Na and Zhang, Yunpeng and Xu, Juan and Li, Yongsheng},
  journal = {Nucleic Acids Research},
  volume = {51},
  number = {D1},
  pages = {D1325--D1332},
  year = {2023},
  doi = {10.1093/nar/gkac922}
}

@article{masopust2026tcellnomenclature,
  title = {Guidelines for T cell nomenclature},
  author = {Masopust, David and Awasthi, Amit and Bosselut, Remy and others},
  journal = {Nature Reviews Immunology},
  volume = {26},
  pages = {298--313},
  year = {2026},
  doi = {10.1038/s41577-025-01238-2}
}

@article{diehl2016cellontology,
  title = {The Cell Ontology 2016: enhanced content, modularization, and ontology interoperability},
  author = {Diehl, Alexander D. and Meehan, Terrence F. and Bradford, Yvonne M. and Brush, Matthew H. and Dahdul, Wasila M. and Dougall, David S. and He, Yongqun and Osumi-Sutherland, David and Ruttenberg, Alan and Sarntivijai, Sirarat and Van Slyke, Ceri E. and Vasilevsky, Nicole A. and Haendel, Melissa A. and Blake, Judith A. and Mungall, Christopher J. and Lewis, Suzanna E.},
  journal = {Journal of Biomedical Semantics},
  volume = {7},
  pages = {44},
  year = {2016},
  doi = {10.1186/s13326-016-0088-7}
}

@article{tan2026cellontology,
  title = {The Cell Ontology in the age of single-cell omics},
  author = {Tan, Shawn Z. K. and Puig-Barbe, Alba and Goutte-Gattat, David and Eastwood, Benjamin S. and Aevermann, Brian D. and others},
  journal = {Scientific Data},
  year = {2026},
  doi = {10.1038/s41597-026-07173-8}
}

@article{czcellxgene2025discover,
  title = {{CZ CELLxGENE} Discover: a single-cell data platform for scalable exploration, analysis and modeling of aggregated data},
  author = {{CZ CELLxGENE Discover Consortium} and Abdulla, Siraj and Aevermann, Brian and Assis, Paulo and Badajoz, Sarah and others},
  journal = {Nucleic Acids Research},
  volume = {53},
  number = {D1},
  pages = {D886--D900},
  year = {2025},
  doi = {10.1093/nar/gkae1142}
}

@article{wang2021onclass,
  title = {Leveraging the Cell Ontology to classify unseen cell types},
  author = {Wang, Sheng and Pisco, Angela Oliveira and McGeever, Aaron and Brbic, Maria and Zitnik, Marinka and Darmanis, Spyros and Leskovec, Jure and Karkanias, Jim and Altman, Russ B.},
  journal = {Nature Communications},
  volume = {12},
  pages = {5556},
  year = {2021},
  doi = {10.1038/s41467-021-25725-x}
}

@article{ergen2024popv,
  title = {Consensus prediction of cell type labels in single-cell data with popV},
  author = {Ergen, Can and Xing, Galen and Xu, Chenling and Kim, Martin and Jayasuriya, Michael and McGeever, Erin and Pisco, Angela Oliveira and Streets, Aaron and others},
  journal = {Nature Genetics},
  volume = {56},
  pages = {2731--2738},
  year = {2024},
  doi = {10.1038/s41588-024-01993-3}
}

@article{lee2020biobert,
  title = {{BioBERT}: a pre-trained biomedical language representation model for biomedical text mining},
  author = {Lee, Jinhyuk and Yoon, Wonjin and Kim, Sungdong and Kim, Donghyeon and Kim, Sunkyu and So, Chan Ho and Kang, Jaewoo},
  journal = {Bioinformatics},
  volume = {36},
  number = {4},
  pages = {1234--1240},
  year = {2020},
  doi = {10.1093/bioinformatics/btz682}
}

@inproceedings{beltagy2019scibert,
  title = {{SciBERT}: A Pretrained Language Model for Scientific Text},
  author = {Beltagy, Iz and Lo, Kyle and Cohan, Arman},
  booktitle = {Proceedings of the 2019 Conference on Empirical Methods in Natural Language Processing and the 9th International Joint Conference on Natural Language Processing},
  pages = {3613--3618},
  year = {2019},
  doi = {10.18653/v1/D19-1371}
}

@article{gu2022pubmedbert,
  title = {Domain-specific language model pretraining for biomedical natural language processing},
  author = {Gu, Yu and Tinn, Robert and Cheng, Hao and Lucas, Michael and Usuyama, Naoto and Liu, Xiaodong and Naumann, Tristan and Gao, Jianfeng and Poon, Hoifung},
  journal = {ACM Transactions on Computing for Healthcare},
  volume = {3},
  number = {1},
  pages = {1--23},
  year = {2022},
  doi = {10.1145/3458754}
}

@article{hou2024gpt4celltype,
  title = {Assessing {GPT}-4 for cell type annotation in single-cell {RNA}-seq analysis},
  author = {Hou, Wenpin and Ji, Zhicheng},
  journal = {Nature Methods},
  volume = {21},
  pages = {1462--1465},
  year = {2024},
  doi = {10.1038/s41592-024-02235-4}
}

@article{wu2025scextract,
  title = {{scExtract}: leveraging large language models for fully automated single-cell {RNA}-seq data annotation and prior-informed multi-dataset integration},
  author = {Wu, Yuxuan and Tang, Fuchou},
  journal = {Genome Biology},
  volume = {26},
  number = {1},
  pages = {174},
  year = {2025},
  doi = {10.1186/s13059-025-03639-x}
}

@article{xie2025cassia,
  title = {{CASSIA}: a multi-agent large language model for automated and interpretable cell annotation},
  author = {Xie, Elliot and Cheng, Lingxin and Shireman, Jack and Cai, Yujia and Liu, Jihua and others},
  journal = {Nature Communications},
  volume = {17},
  pages = {389},
  year = {2026},
  doi = {10.1038/s41467-025-67084-x}
}

@article{robertson2009BM25,
  title={The probabilistic relevance framework: BM25 and beyond},
  author={Robertson, Stephen and Zaragoza, Hugo and others},
  journal={Foundations and Trends{\textregistered} in Information Retrieval},
  year={2009},
}

@article{ristad1998editDist,
  title={Learning string-edit distance},
  author={Ristad, Eric Sven and Yianilos, Peter N},
  journal={TPAMI},
  year={1998},
}

@inproceedings{liu2021SapBERT,
  title={Self-Alignment Pretraining for Biomedical Entity Representations},
  author={Liu, Fangyu and Shareghi, Ehsan and Meng, Zaiqiao and Basaldella, Marco and Collier, Nigel},
  booktitle={NAACL},
  year={2021}
}

@misc{zooma,
  author       = {{EMBL-EBI}},
  title        = {{ZOOMA}: Ontology annotation tool},
  howpublished = {\url{https://www.ebi.ac.uk/spot/zooma/}},
  year         = {Accessed 2026} 
}

@article{bioportal,
    author = {Noy, Natalya F. and Shah, Nigam H. and Whetzel, Patricia L. and Dai, Benjamin and Dorf, Michael and Griffith, Nicholas and Jonquet, Clement and Rubin, Daniel L. and Storey, Margaret-Anne and Chute, Christopher G. and Musen, Mark A.},
    title = {BioPortal: ontologies and integrated data resources at the click of a mouse},
    journal = {Nucleic Acids Research},
    volume = {37},
    number = {suppl\_2},
    pages = {W170-W173},
    year = {2009},
    month = {07},
    doi = {10.1093/nar/gkp440},
}

@article{kuhn1955hungarian,
  title     = {The Hungarian method for the assignment problem},
  author    = {Kuhn, Harold W.},
  journal   = {Naval Research Logistics Quarterly},
  volume    = {2},
  number    = {1-2},
  pages     = {83--97},
  year      = {1955},
  doi       = {10.1002/nav.3800020109},
}

@article{vinh2010information,
  title = {Information Theoretic Measures for Clusterings Comparison: Variants, Properties, Normalization and Correction for Chance},
  author = {Vinh, Nguyen Xuan and Epps, Julien and Bailey, James},
  journal = {Journal of Machine Learning Research},
  volume = {11},
  number = {95},
  pages = {2837--2854},
  year = {2010},
  url = {https://www.jmlr.org/papers/v11/vinh10a.html}
}

@article{youngblut2025scbasecount,
  title = {{scBaseCount}: an {AI} agent-curated, uniformly processed, and autonomously updated single cell data repository},
  author = {Youngblut, Nicholas D. and Carpenter, Christopher and Nayebnazar, Arshia and Adduri, Abhinav and Shah, Rohan and Ricci-Tam, Chiara and Prashar, Jaanak and Ilango, Rajesh and Teyssier, Noam and Konermann, Silvana and Hsu, Patrick D. and Dobin, Alexander and Burke, Dave P. and Goodarzi, Hani and Roohani, Yusuf H.},
  journal = {bioRxiv},
  year = {2025},
  doi = {10.1101/2025.02.27.640494}
}

@article{cao2019organogenesis,
  title = {The single-cell transcriptional landscape of mammalian organogenesis},
  author = {Cao, Junyue and Spielmann, Malte and Qiu, Xiaojie and others},
  journal = {Nature},
  volume = {566},
  number = {7745},
  pages = {496--502},
  year = {2019},
  doi = {10.1038/s41586-019-0969-x}
}

@article{lake2023kidneyatlas,
  title = {An atlas of healthy and injured cell states and niches in the human kidney},
  author = {Lake, Blue B. and Menon, Rajasree and Winfree, Seth and others},
  journal = {Nature},
  volume = {619},
  number = {7970},
  pages = {585--594},
  year = {2023},
  doi = {10.1038/s41586-023-05769-3}
}

@article{wilk2020immuneatlas,
  title = {A single-cell atlas of the peripheral immune response in patients with severe {COVID-19}},
  author = {Wilk, Aaron J. and Rustagi, Arjun and Zhao, Nancy Q. and others},
  journal = {Nature Medicine},
  volume = {26},
  number = {7},
  pages = {1070--1076},
  year = {2020},
  doi = {10.1038/s41591-020-0944-y}
}

@article{regev2017humancellatlas,
  title = {The {Human Cell Atlas}},
  author = {Regev, Aviv and Teichmann, Sarah A. and Lander, Eric S. and others},
  journal = {{eLife}},
  volume = {6},
  pages = {e27041},
  year = {2017},
  doi = {10.7554/eLife.27041}
}

@article{han2018mousecellatlas,
  title = {Mapping the {Mouse Cell Atlas} by {Microwell-Seq}},
  author = {Han, Xiaoping and Wang, Renying and Zhou, Yincong and others},
  journal = {Cell},
  volume = {172},
  number = {5},
  pages = {1091--1107.e17},
  year = {2018},
  doi = {10.1016/j.cell.2018.02.001}
}

@article{tabulamuris2018atlas,
  title = {Single-cell transcriptomics of 20 mouse organs creates a {Tabula Muris}},
  author = {{Tabula Muris Consortium}},
  journal = {Nature},
  volume = {562},
  number = {7727},
  pages = {367--372},
  year = {2018},
  doi = {10.1038/s41586-018-0590-4}
}

@article{wu2021breastatlas,
  title = {A single-cell and spatially resolved atlas of human breast cancers},
  author = {Wu, Sunny Z. and Al-Eryani, Ghamdan and Roden, Daniel Lee and others},
  journal = {Nature Genetics},
  volume = {53},
  number = {9},
  pages = {1334--1347},
  year = {2021},
  doi = {10.1038/s41588-021-00911-1}
}

@article{liu2021nknkt,
  title = {{NK} and {NKT} cells have distinct properties and functions in cancer},
  author = {Liu, Xia and Li, Lingyun and Si, Fusheng and Huang, Lan and Zhao, Yangjing and Zhang, Chenchen and Hoft, Daniel F. and Peng, Guangyong},
  journal = {Oncogene},
  volume = {40},
  number = {27},
  pages = {4521--4537},
  year = {2021},
  doi = {10.1038/s41388-021-01880-9}
}

@article{guerreromurillo2024cartmultiomics,
  title = {Integrative single-cell multi-omics of {CD19-CAR}\textsuperscript{pos} and {CAR}\textsuperscript{neg} {T} cells suggest drivers of immunotherapy response in {B}-cell neoplasias},
  author = {Guerrero-Murillo, Mercedes and Rill-Hinarejos, Aina and Trincado, Juan L. and others},
  journal = {bioRxiv},
  year = {2024},
  doi = {10.1101/2024.01.23.576878}
}

@article{chu2023pancancertcellatlas,
  title = {Pan-cancer {T} cell atlas links a cellular stress response state to immunotherapy resistance},
  author = {Chu, Yanshuo and Dai, Enyu and Li, Yating and others},
  journal = {Nature Medicine},
  volume = {29},
  number = {6},
  pages = {1550--1562},
  year = {2023},
  doi = {10.1038/s41591-023-02371-y}
}

@article{han2022follicularlymphoma,
  title = {Follicular Lymphoma Microenvironment Characteristics Associated with Tumor Cell Mutations and {MHC} Class {II} Expression},
  author = {Han, Guangchun and Deng, Qing and Marques-Piubelli, Mario L. and others},
  journal = {Blood Cancer Discovery},
  volume = {3},
  number = {5},
  pages = {428--443},
  year = {2022},
  doi = {10.1158/2643-3230.BCD-21-0075}
}

@article{kurkalang2023oralcancer,
  title = {Single-cell transcriptomic analysis of gingivo-buccal oral cancer reveals two dominant cellular programs},
  author = {Kurkalang, Sillarine and Roy, Sumitava and Acharya, Arunima and others},
  journal = {Cancer Science},
  volume = {114},
  number = {12},
  pages = {4732--4746},
  year = {2023},
  doi = {10.1111/cas.15979}
}

@article{vazquezgarcia2022ovarian,
  title = {Ovarian cancer mutational processes drive site-specific immune evasion},
  author = {V{\'a}zquez-Garc{\'i}a, Ignacio and Uhlitz, Florian and Ceglia, Nicholas and others},
  journal = {Nature},
  volume = {612},
  number = {7941},
  pages = {778--786},
  year = {2022},
  doi = {10.1038/s41586-022-05496-1}
}

@article{wu2020tnbcstroma,
  title = {Stromal cell diversity associated with immune evasion in human triple-negative breast cancer},
  author = {Wu, Sunny Z. and Roden, Daniel L. and Wang, Chenfei and others},
  journal = {The EMBO Journal},
  volume = {39},
  number = {19},
  year = {2020},
  doi = {10.15252/embj.2019104063}
}

@article{osumi2021cell,
  title={Cell type ontologies of the Human Cell Atlas},
  author={Osumi-Sutherland, David and Xu, Chuan and Keays, Maria and Levine, Adam P and Kharchenko, Peter V and Regev, Aviv and Lein, Ed and Teichmann, Sarah A},
  journal={Nature cell biology},
  volume={23},
  number={11},
  pages={1129--1135},
  year={2021},
  publisher={Nature Publishing Group UK London}
}

@article{hu2023cellmarker,
  title={CellMarker 2.0: an updated database of manually curated cell markers in human/mouse and web tools based on scRNA-seq data},
  author={Hu, Congxue and Li, Tengyue and Xu, Yingqi and Zhang, Xinxin and Li, Feng and Bai, Jing and Chen, Jing and Jiang, Wenqi and Yang, Kaiyue and Ou, Qi and others},
  journal={Nucleic acids research},
  volume={51},
  number={D1},
  pages={D870--D876},
  year={2023},
  publisher={Oxford University Press}
}

@article{franzen2019panglaodb,
  title={PanglaoDB: a web server for exploration of mouse and human single-cell RNA sequencing data},
  author={Franz{\'e}n, Oscar and Gan, Li-Ming and Bj{\"o}rkegren, Johan LM},
  journal={Database},
  volume={2019},
  pages={baz046},
  year={2019},
  publisher={Oxford University Press}
}

@article{yuste2020community,
  title={A community-based transcriptomics classification and nomenclature of neocortical cell types},
  author={Yuste, Rafael and Hawrylycz, Michael and Aalling, Nadia and Aguilar-Valles, Argel and Arendt, Detlev and Arma{\~n}anzas, Ruben and Ascoli, Giorgio A and Bielza, Concha and Bokharaie, Vahid and Bergmann, Tobias Borgtoft and others},
  journal={Nature neuroscience},
  volume={23},
  number={12},
  pages={1456--1468},
  year={2020},
  publisher={Nature Publishing Group US New York}
}

@article{spits2013innate,
  title={Innate lymphoid cells—a proposal for uniform nomenclature},
  author={Spits, Hergen and Artis, David and Colonna, Marco and Diefenbach, Andreas and Di Santo, James P and Eberl, Gerard and Koyasu, Shigeo and Locksley, Richard M and McKenzie, Andrew NJ and Mebius, Reina E and others},
  journal={Nature reviews immunology},
  volume={13},
  number={2},
  pages={145--149},
  year={2013},
  publisher={Nature Publishing Group UK London}
}

@article{bard2005ontology,
  title={An ontology for cell types},
  author={Bard, Jonathan and Rhee, Seung Y and Ashburner, Michael},
  journal={Genome biology},
  volume={6},
  number={2},
  pages={R21},
  year={2005},
  publisher={Springer}
}

@article{saunders2018molecular,
  title={Molecular diversity and specializations among the cells of the adult mouse brain},
  author={Saunders, Arpiar and Macosko, Evan Z and Wysoker, Alec and Goldman, Melissa and Krienen, Fenna M and de Rivera, Heather and Bien, Elizabeth and Baum, Matthew and Bortolin, Laura and Wang, Shuyu and others},
  journal={Cell},
  volume={174},
  number={4},
  pages={1015--1030},
  year={2018},
  publisher={Elsevier}
}

@inproceedings{ma2025scpankd,
  title={ScPanKD: Distilling Pan-Cancer Knowledge for Enhanced T Cell Subtypes Annotation in Single-Cell Transcriptomics Data},
  author={Ma, Wenjing and Yu, Xiaoqing and Lu, Jiaying and Zhang, Jing and Zhou, Xiang and Wang, Xuefeng},
  booktitle={2025 IEEE International Conference on Bioinformatics and Biomedicine (BIBM)},
  pages={532--537},
  year={2025},
  organization={IEEE}
}

@misc{xiaomi2026mimov25,
      title={Full-Pipeline Inference Optimization for MiMo-V2.5 Series: Pushing Hybrid SWA Efficiency to the Limit}, 
      author={Xiaomi MiMo Team and Anqi Liu and Aoxin Ma and Bo Chen and Bo Yang and Chen Wang and Chen Zhang and Chengda Tang and Chengwei Wang and Chiheng Lou and Depeng Yan and Fuli Luo and Gang Wang and Hailin Zhang and Jiale Sun and Kang Zhou and Rui Huang and Shaohui Liu and Shen Huang and Shijie Cao and Shuaishuai Fan and Tianling Zhou and Xiangwei Deng and Xueyang Xie and Xuli Wang and Yingchun Lai and Yu Yang and Yuan Zhang and Zhen Tang and Zhonghua Deng and Zihan Jiang},
      year={2026},
      eprint={2607.13095},
      archivePrefix={arXiv},
      primaryClass={cs.AR},
      url={https://arxiv.org/abs/2607.13095}, 
}

@article{xu2026deepseekv4,
  title={Deepseek-v4: Towards highly efficient million-token context intelligence},
  author={Xu, Anyi and Lin, Bangcai and Xue, Bing and Wang, Bingxuan and Xu, Bingzheng and Wu, Bochao and Zhang, Bowei and Lin, Chaofan and Dong, Chen and Ling, Chenchen and others},
  journal={arXiv preprint arXiv:2606.19348},
  year={2026}
}

@article{zeng2026glm5,
  title={Glm-5: from vibe coding to agentic engineering},
  author={Zeng, Aohan and Lv, Xin and Hou, Zhenyu and Du, Zhengxiao and Zheng, Qinkai and Chen, Bin and Yin, Da and Ge, Chendi and Huang, Chenghua and Xie, Chengxing and others},
  journal={arXiv preprint arXiv:2602.15763},
  year={2026}
}

@misc{deepmind2026gemini35flash,
  author = {{Google DeepMind}},
  title = {Gemini 3.5 Flash Model Card},
  year = {2026},
  month = may,
  institution = {Google DeepMind},
  url = {https://deepmind.google/models/model-cards/gemini-3-5-flash/},
  note = {Released May 19, 2026}
}

@misc{anthropic_2026_claude_opus_48,
  author = {{Anthropic}},
  title = {System Card: Claude Opus 4.8},
  year = {2026},
  month = may,
  institution = {Anthropic},
  url = {https://www.anthropic.com/claude-opus-4-8-system-card},
  note = {Released May 28, 2026}
}

@techreport{openai_2026_gpt55,
  author = {{OpenAI}},
  title = {GPT-5.5 System Card},
  year = {2026},
  month = apr,
  institution = {OpenAI},
  url = {https://deploymentsafety.openai.com/gpt-5-5/gpt-5-5.pdf},
  note = {Released April 23, 2026}
}

\clearpage
\appendix
\renewcommand{\thetable}{S\arabic{table}}
\setcounter{table}{0}
\section{Literature Retrieval and Single-Cell Data Acquisition}
\label{sec:supp-literature-data}

To evaluate the consistency of T-cell nomenclature across physiological and disease contexts, we constructed a benchmark linking scientific publications to publicly available single-cell datasets. Inspired by the large-scale single-cell data discovery and curation workflow of scBaseCount~\cite{youngblut2025scbasecount}, we focused our data collection on four study contexts: Healthy, Cancer, Infectious, and Inflammatory.

We first used a data-first strategy to construct the Healthy reference set. Starting from 361 publications indexed in CZ CELLxGENE and linked to DOIs, we filtered studies by organism, assay modality, health status, and the presence of T-cell annotations, yielding 65 candidate papers. Of these, 21 passed full-text screening, and the 20 most highly cited studies were retained.

For the three disease contexts of Cancer, Infectious, and Inflammatory, we searched the Gene Expression Omnibus (GEO) using combinations of disease-specific, T-cell, and single-cell transcriptomics terms. We further screened the retrieved studies for paper relevance and duplicated DOIs, and matched eligible studies to their corresponding H5AD datasets in CZ CELLxGENE. Together with the Healthy reference set, 66 papers proceeded to manual review.

We then manually reviewed the 66 in-scope papers and their matched H5AD data. We retained a study only when its H5AD data corresponded to the T-cell analysis reported in the paper and included an appropriate author-provided annotation field. We additionally required a verifiable correspondence between the paper-reported labels and the dataset annotations. In total, 44 studies met these criteria and formed the final benchmark, comprising 12 Healthy, 13 Cancer, 13 Infectious, and 6 Inflammatory studies.

\section{Information Extraction Evaluation Set}
\label{sec:supp-stage1-evaluation-set}

Stage~1 was evaluated on the seven papers listed in Table~\ref{tab:supp-stage1-paper-inventory}, with four domain experts independently annotating each paper for its reported T-cell population labels and supporting evidence. This evaluation set is separate from the 44-study benchmark used for nomenclature alignment, although four papers occur in both.

\newtcblisting{promptlisting}[1]{
  promptbox,
  colback=black!1!white,
  colframe=black!35,
  colbacktitle=black!8,
  coltitle=black,
  title={#1},
  listing only,
  listing options={
    basicstyle=\ttfamily\scriptsize,
    breaklines=true,
    columns=fullflexible,
    keepspaces=true,
    showstringspaces=false
  }
}

\section{Prompt Templates Used in \model}
\label{sec:supp-prompts}

The following templates present the instruction text used in the workflow. Angle-bracket fields denote paper- or case-specific content substituted at runtime and were not sent literally; implemen\-tation-only file labels are omitted.

\subsection{Search Agent Prompts}

\subsubsection{Paper-Level Screening Prompt}

\begin{promptlisting}{Paper-level screening}
[System]
Return valid JSON only.

[User]
You are a paper-level screening classifier.

Your task is to decide whether the input paper contains extractable, paper-specific T-cell type, subtype, state, or population content.

A paper should return true only when ALL of the following conditions are satisfied:

1. The paper analyzes T cells / T lymphocytes as part of its own biological results.

2. The paper describes one or more specific T-cell types, subtypes, states, populations, clusters, or functionally distinct T-cell groups.

3. The T-cell categories are described with paper-specific biological information, such as marker genes, marker proteins, functional programs, phenotypic states, activation or exhaustion features, tissue or tumor context, spatial localization, abundance changes, clinical associations, or experimental findings.

4. The T-cell information is specific enough that a structured extractor could list the T-cell categories and attach supporting evidence from the paper text.

Return exactly one JSON object with exactly one key:

{"paper_passes_screening": true}

or

{"paper_passes_screening": false}

Full-paper Markdown:

<paper_markdown>
<FULL-PAPER MARKDOWN>
</paper_markdown>
\end{promptlisting}

\subsubsection{Single-Cell Data-Clue Extraction Prompt}

\begin{promptlisting}{Single-cell data-clue extraction}
[System]
Return valid JSON only.

[User]
You are a Stage 0 scRNA-seq h5ad data clue extractor.

Your task is to read the full-paper Markdown and extract only the clues needed for code to verify whether this paper has usable scRNA-seq h5ad data through CELLxGENE.

Do not decide whether the paper passes Stage 0. Only extract verification clues.

Extract paper identity when available. These fields are verification clues: code will use them to search CELLxGENE metadata even when the paper does not explicitly mention CELLxGENE.
- title
- DOI
- PMID
- journal
- year

Extract general scRNA-seq evidence from the paper whenever the paper describes scRNA-seq or single-cell transcriptomic RNA expression data, even if the paper does not explicitly mention CELLxGENE:
- assay hint, always "scRNA-seq"
- relation to this paper
- short evidence text copied or closely paraphrased from the paper

Do not extract GEO, SRA, ENA, EGA, FASTQ, count matrix, or other raw-data repository clues as passing evidence. Raw data availability is not enough for this Stage 0 condition.

Extract CELLxGENE clues only when the CELLxGENE collection or dataset is described as containing scRNA-seq or single-cell transcriptomic RNA expression data:
- CELLxGENE URLs
- collection names
- dataset names
- collection DOI or dataset DOI
- assay hint, always "scRNA-seq"
- relation to this paper

If a paper is multi-modal, extract the clue only when it explicitly includes a scRNA-seq or single-cell RNA expression component. Do not extract clues for records that are only snRNA-seq, spatial transcriptomics, CITE-seq, bulk RNA-seq, ATAC-seq, proteomics, imaging, or other non-scRNA-seq assays.

Use this allowed value for assay_hint:
- scRNA-seq

Use these allowed values for relation_to_paper:
- generated_by_this_paper
- reused_external_data
- unclear

Use "generated_by_this_paper" when the paper appears to have generated or directly analyzed the dataset.
Use "reused_external_data" when the paper only reuses an external scRNA-seq dataset.
Use "unclear" only when the relation cannot be determined from the Markdown.

If no scRNA-seq evidence is found anywhere in the paper, return "scrnaseq_evidence": [].
If no scRNA-seq CELLxGENE clues are found, return "cellxgene_clues": [].
Return "mentioned_repositories" only for CELLxGENE mentions tied to scRNA-seq h5ad data.
If no scRNA-seq CELLxGENE repository is mentioned, return "mentioned_repositories": [].

Return exactly one JSON object using this schema:

{
  "paper_identity": {
    "title": null,
    "doi": null,
    "pmid": null,
    "journal": null,
    "year": null
  },
  "scrnaseq_evidence": [
    {
      "assay_hint": "scRNA-seq",
      "relation_to_paper": "generated_by_this_paper",
      "evidence_text": "The scRNA-seq data from this study were deposited in repository X."
    }
  ],
  "cellxgene_clues": [
    {
      "url": null,
      "collection_name": null,
      "dataset_name": null,
      "doi": null,
      "assay_hint": "scRNA-seq",
      "relation_to_paper": "generated_by_this_paper"
    }
  ],
  "mentioned_repositories": ["CELLxGENE"]
}

Full-paper Markdown:

<paper_markdown>
<FULL-PAPER MARKDOWN>
</paper_markdown>
\end{promptlisting}

\subsection{Information Extraction Agent Prompt}

\subsubsection{T-Cell Label Extraction Prompt}

\begin{promptlisting}{T-cell label extraction}
[System]
Return valid JSON only. Do not include markdown fences.

[User]
You are an immunology literature extraction assistant. Read the provided PDF and extract T cell subtypes / T cell states / T cell populations that are explicitly supported by evidence in the article.

Core extraction principle:
The output unit is the biological T cell subtype/state/population, not the individual cluster.

Extraction criteria:
Each final item must satisfy all of the following criteria:

1. It is a T cell-related subtype, state, or population.
2. It explicitly appears in the paper or is explicitly annotated/defined by the authors.
3. It has supporting evidence: gene expression or descriptive words.
4. If both a broad parent label and a more specific biological subtype/state are present, keep the more specific supported subtype/state/population as the final item.

Original label wording rule:
- original_label must be a complete, self-contained biological T cell subtype/state/population label.
- Keep original_label as close as possible to the paper's original wording.
- Prefer a complete biological label over a local shorthand, abbreviation-only label, cluster-only label, or subcluster-only label.
- A local shorthand label is incomplete if it only contains an abbreviation, cluster ID, subcluster ID, or state ID, but its biological meaning is defined elsewhere in the paper.
- If the paper defines a local shorthand, abbreviation, cluster ID, or subcluster ID under a broader biological label, use the broader biological label to construct a complete original_label, and preserve the shorthand/ID in parentheses.
- Use only biological meanings stated or clearly defined in the paper.
- Do not expand abbreviations using external immunology knowledge.
- Do not output bare local shorthand labels as final original_label when the paper provides the biological meaning elsewhere.
- When merging several equivalent mentions, choose the clearest representative wording from the paper.
- If multiple cluster IDs belong to the same biological subtype/state/population, include the merged cluster IDs in original_label when needed to preserve the paper's context.
- Do not normalize the label into an external ontology or rewrite it into a standard name not used by the paper.

Merge rule:
- Merge mentions that refer to the same biological T cell subtype/state/population.
- Do not create separate final items only because the same biological subtype/state/population appears in multiple clusters, samples, figures, sections, or analysis views.
- Keep cluster/subcluster/state IDs inside original_label only when they are part of the paper's wording or needed to preserve context.
- If two labels have different biological meanings in the paper, keep them as separate final items.

Evidence rule:
- Each evidence item must be gene expression or descriptive words.
- Keep every evidence item strictly in the paper's original wording.
- Do not rewrite, paraphrase, translate, summarize, or normalize evidence.
- Evidence should be copied as short original phrases from the article.
- The same original_label may have multiple evidence items.
- If the same cell subtype/state/population is supported by multiple distinct evidence phrases, include all relevant evidence phrases in the same evidence array.

Exclusion rules:
- Do not extract non-T cell types.
- Do not extract labels that appear without gene expression or descriptive words.
- Do not invent subtypes from general immunology knowledge; use only information from the paper.
- Do not treat marker genes themselves as cell subtypes.
- Do not include T cell names that are mentioned only in background, introduction, references, or general discussion without being part of the paper's own results or annotations.

Pre-output label check:
Before returning JSON, review every original_label.

For each item:
1. Ask whether the label is only a local shorthand, abbreviation, cluster ID, subcluster ID, or state ID.
2. If yes, look for the biological label that the paper uses to define that shorthand.
3. If a biological label is available in the paper, rewrite original_label as a complete, self-contained, paper-faithful biological label that includes the shorthand/ID in parentheses.
4. If no biological label is available in the paper, keep the paper's local shorthand label unchanged.
5. Never use external knowledge to expand or rename the label.

Output requirements:
Return valid JSON only. Do not output markdown or explanations.

JSON schema:
{
  "title": "",
  "items": [
    {
      "original_label": "A complete, self-contained, paper-faithful biological T cell subtype/state/population label. Include paper abbreviation, cluster ID, subcluster ID, or state ID in parentheses when needed.",
      "evidence": [
        "Exact original gene expression or descriptive words from the paper",
        "Another exact original evidence phrase for the same biological cell subtype/state/population, if available"
      ]
    }
  ]
}

Stage 0 paper markdown input:
Use only the supplied full-paper markdown below.

SUPPLIED TEXT START
<FULL-PAPER MARKDOWN>
SUPPLIED TEXT END
\end{promptlisting}

\subsection{Mapping Agent Prompt}

\subsubsection{Nomenclature-Guided Mapping Prompt}

\begin{promptlisting}{Nomenclature-guided mapping}
[User]
Use the reference to complete structured T cell nomenclature slot filling for the following T cell cases.

Return only valid JSON. Do not include explanations.
Return one output object per input case.
Do not copy original_label or evidence into the output.
Do not return mapped_label. The final mapped_label will be derived by code from mapped_slots.

For each supported slot, return:
- module: the final naming module inferred from the reference.
- semantic_value: a concise snake_case semantic key describing the biological meaning of that module, inferred from the reference.

If the evidence does not support a slot, return:
{"module": null, "semantic_value": null}

Do not invent unsupported biological properties.
Do not fill a slot only because it is familiar from the label name.
Prefer less specific correct slots over more specific unsupported slots.
Use the same semantic_value string consistently for the same biological meaning across cases.

REFERENCE:
<<<
<T-CELL NOMENCLATURE REFERENCE>
>>>

CASES:
<MAPPING CASES JSON>

Output schema:
[
  {
    "case_id": 0,
    "mapped_slots": {
      "lineage": {
        "module": "CD4+",
        "semantic_value": "cd4_conventional"
      },
      "function": {
        "module": null,
        "semantic_value": null
      },
      "migration": {
        "module": null,
        "semantic_value": null
      },
      "migration_subscript": {
        "module": null,
        "semantic_value": null
      },
      "differentiation": {
        "module": null,
        "semantic_value": null
      },
      "differentiation_subscript": {
        "module": null,
        "semantic_value": null
      },
      "antigen": {
        "module": null,
        "semantic_value": null
      }
    }
  }
]
\end{promptlisting}

\subsubsection{T-Cell Nomenclature Reference}

The following reference text was supplied to the Mapping Agent through the reference field shown above.

\begin{tcblisting}{
  promptbox,
  colback=black!1!white,
  colframe=black!35,
  listing only,
  listing options={
    basicstyle=\ttfamily\scriptsize,
    breaklines=true,
    columns=fullflexible,
    keepspaces=true,
    showstringspaces=false
  }
}
# T Cell Type Nomenclature Rule

## 1. Core Principle

Use a **modular T cell name** when a paper describes a T cell type, state, subset, or population.

The name should encode only the biological properties that are actually stated or reasonably supported by the paper:

`[lineage] T[function][migration][migration-subscript][differentiation-state][state-subscript][antigen-status]`

If a property is not stated or not measured, do **not** infer it from a familiar subset label.

## 2. Core Naming Syntax

Use this order:

1. Lineage
2. `T` root
3. Functional module
4. Migration module
5. Migration subscript
6. Differentiation-state module
7. Differentiation-state subscript
8. Antigen-status suffix
9. No terminal standalone word `cell` after the modular code

For plain-text LLM input, serialize subscripts inline in the canonical order.

Do not reorder modules for readability. Keep function before migration, migration before differentiation state, and antigen status last.

## 3. Lineage Module

Lineage is the broad, relatively stable cell identity.

Allowed lineage labels include:

| Module | Meaning |
|---|---|
| `CD4+` | Conventional CD4-positive T cell |
| `CD8+` | Conventional CD8-positive T cell |
| `gamma-delta+` or `gamma-delta T` | gamma-delta T cell lineage |
| `NKT` / `iNKT` | Natural killer T cell / invariant NKT cell |
| `MAIT` | Mucosal-associated invariant T cell |
| `CD8aa+ TCRab` | CD8 alpha-alpha-positive TCR alpha-beta T cell |
| `DN T` | Double-negative T cell |
| `DP T` | Extrathymic double-positive T cell |

Rule:

- Treat stable lineage identities such as `CD4+`, `CD8+`, `gamma-delta T`, `MAIT`, and `iNKT` as cell-type-like labels.
- Do not convert transient activation, memory, exhaustion, or tissue-location descriptions into lineage labels.

## 4. Functional Module

Functional modules follow the `T` root and precede migration/state modules.

Common functional modules:

| Module | Meaning |
|---|---|
| `TH1` | CD4 T cell biased toward IFN-gamma-associated type 1 helper function |
| `TH2` | CD4 T cell biased toward IL-4/IL-5/IL-13-associated type 2 helper function |
| `TH9` | CD4 T cell biased toward IL-9-associated helper function |
| `TH17` | CD4 T cell biased toward IL-17-associated helper function |
| `TFH` | Follicular helper T cell function |
| `Treg` | Regulatory T cell function |
| `tTreg` | Thymus-derived regulatory T cell |
| `pTreg` | Peripherally derived regulatory T cell |
| `eTreg` | Effector regulatory T cell |
| `iTreg` | In vitro-induced regulatory T cell |
| `TFR` | Follicular regulatory T cell |
| `TCTL` | Cytotoxic T cell function |
| `TC2`, `TC17`, `TC22` | CD8 T cells with helper-like cytokine/function programs |

Rules:

- Use functional labels only when the paper states functional identity, transcription-factor evidence, cytokine evidence, or an explicit author claim.
- Keep entrenched `TH` and `Treg` labels; do not erase them merely because lineage is already `CD4+`.
- Use `TCTL` for cytotoxic function when cytotoxicity is the intended property. Do not use `C` alone to mean cytotoxic, because `C` is historically ambiguous.
- Avoid introducing `C` or `E` as new modular shorthand for central/effector migration categories, because they are easily confused with `TCM` and `TEM`.

## 5. Migration Module

Migration is encoded separately from memory, activation, exhaustion, or function.

| Module | Meaning | Typical evidence/proxy |
|---|---|---|
| `S` | Can migrate from blood into uninflamed secondary lymphoid organs | CD62L+ and/or CCR7+ |
| `D` | Disseminated; does not tend to enter uninflamed secondary lymphoid organs from blood | CD62L- and/or CCR7- |
| `U` | Migration properties unknown | No migration assay or no homing-receptor evidence |

Migration subscripts:

| Subscript | Meaning |
|---|---|
| `B` | Isolated from blood; no further migration claim |
| `R` | Resident; parked within an organ or restricted vascular compartment |
| `W` | Widespread; recirculates through non-lymphoid tissues |

Rules:

- Use `S`, `D`, or `U` only for migration behavior or its accepted proxy markers.
- Use `D` for CD62L- and/or CCR7- cells when the paper only supports failure to enter uninflamed SLOs.
- Do not call a cell `TRM`, resident, or `R` merely because it was isolated from a tissue. Tissue source alone is not residence.
- If a cell is from blood and no further migration is known, use `B`.
- If migration is unknown but blood origin is known, use `UB`, not `DB`.

## 6. Differentiation-State Module

Differentiation state follows migration.

| Module | Meaning |
|---|---|
| `N` | Naive |
| `A` | Activated |
| `M` | Memory |
| `X` | Exhausted |
| `G` | Anergic |

State subscripts:

| Subscript | Meaning |
|---|---|
| `p` | Progenitor / precursor |
| `t` | Terminal |

Rules:

- Use `N` for naive T cells.
- Use `A` for recently activated T cells.
- Use `M` for memory T cells.
- Use `X` for exhausted T cells.
- Use `G` for anergic T cells.
- Use `p` or `t` only when progenitor/precursor or terminal status is stated or supported.
- Map current labels as follows when evidence supports the mapping:
  - `TPEX` -> `TXp`
  - terminal exhausted T cell / `TEX-term` -> `TXt`
  - stem cell memory T cell / `TSCM` -> `TMp`
  - memory precursor effector cell / `MPEC` -> `TAp`
  - short-lived effector cell / `SLEC` -> `TAt`
- Do not use `effector` as a precise state if the paper only says effector-like molecules are expressed. Prefer `A`, `M`, or leave the state unspecified depending on evidence.

## 7. Antigen-Status Suffix

Antigen status is optional and attaches after the differentiation-state module.

| Suffix | Meaning |
|---|---|
| `+` | Cognate antigen is claimed to persist in the organism |
| `0` | Cognate antigen is claimed cleared or irrelevant to the described cell |
| no suffix | Antigen status not stated or not needed |

Rules:

- Use `+` only for explicit or strongly supported persistent-antigen contexts, such as chronic infection or tumor antigen persistence.
- Use `0` only when antigen clearance or irrelevance is stated or reasonably claimed.
- Do not infer antigen status from `memory`, `exhausted`, `TRM`, or tissue location alone.
- Plain `M` does not imply antigen cleared.

## 8. Recommended Combination Order

Canonical order:

`[lineage] T[function][S/D/U][B/R/W][N/A/M/X/G][p/t][+/0]`

## 9. Content to Preserve as Evidence/Context, Not Force Into the Name

Keep the following outside the mapped label unless a defined module exists and the paper makes the relevant claim:

- Tissue or anatomical source: `lung`, `liver`, `blood`, `tumor`, `lymph node`
- Species: human, mouse, non-human primate
- Disease or perturbation context
- Antigen specificity, clone, tetramer identity, or TCR specificity
- Marker evidence used to justify a module
- Cytokines, transcription factors, and effector molecules when they are evidence rather than the intended functional label
- Proliferation potential
- Differentiation potential beyond `p` or `t`
- Longevity
- Developmental plasticity
- Epigenetic profile
- Assay type or gating strategy
- Uncertainty notes such as `claimed`, `not measured`, `proxy only`

## 10. Boundaries: Do Not Over-Standardize

Do not force a familiar subset label into a stronger modular claim than the paper supports.

Rules:

- `TEM` does not automatically mean blood-derived, tissue-recirculating, resident, terminal, cytotoxic, or antigen-cleared.
- `TRM` requires a residence claim; tissue isolation alone is insufficient.
- `TEMRA` is human-specific and should not be projected onto mouse data.
- `Effector T cell` is semantically broad. Map to `A`, `At`, cytotoxic function, or leave the state unspecified if the paper does not give enough evidence.
- `Memory T cell` does not itself imply antigen clearance.
- `Exhausted T cell` usually implies chronic antigen context, but still use `+` only when persistence is claimed or contextually supported.
- Marker-only descriptions should be converted only as far as the markers support. If no claim is made, keep markers in evidence/context.
- If migration markers are absent, use `U` only when the naming task requires migration to be represented; otherwise omit migration.
- If activation recency is unknown, do not add `A`.
- If residence, recirculation, antigen persistence, or terminal/progenitor status is not stated, omit those modules.

## 11. examples

Use these source-paper examples as examples of how method-defined criteria support modular names.

### Example 1

Results text:

```text
We compared populations of naive CD4+ T cells (CD4+ TN) with activated CD4+ T cells (CD4+ TA) and memory CD4+ T cells (CD4+ TM).
```

Methods criteria:

```text
CD4+ TN were CD44low, CD62L+. CD4+ TA were CD44hi, CD62L-, and known to have been stimulated by antigen within the last week. CD4+ TM were CD44hi and the specific pathogen is thought to have been cleared at least 1 month before analysis.
```

### Example 2

Results text:

```text
We studied influenza virus-specific disseminated resident memory CD8+ T cells (CD8+ TDRM) in the mediastinal lymph node.
```

Methods criteria:

```text
We used a CD69+, CD62L- flow cytometry phenotype to define CD8+ TDRM. Influenza-specific CD69+, CD62L- CD8+ T cells were previously shown to be resident based on minimal equilibration in parabionts.
```

### Example 3

Results text:

```text
We profiled exhausted progenitor CD8+ T cells (TXp) by single-cell RNA sequencing.
```

Methods criteria:

```text
All CD8+ T cells within a cluster that was distinguished by elevated expression of TCF7, PDCD1 and CXCR5 were defined as TXp.
```

## 12. Decision Rule

When normalizing T cell population names:

1. Preserve the original mention.
2. Extract only stated or supported biological properties.
3. Build the modular name in canonical order.
4. Store unsupported implications as uncertainty, not as modules.
5. Omit a final generic `cell` suffix from modular mapped labels.
6. Prefer a less specific correct name over a more specific over-inferred name.
\end{tcblisting}

\subsection{Judge Agent Prompt}

\subsubsection{Slot-Level Adjudication Prompt}

\begin{promptlisting}{Slot-level adjudication}
[System]
Return only valid JSON matching the requested schema.
Do not include markdown fences or any text outside the JSON object.

[User]
Judge whether every structured slot in the candidate T cell mapping is valid under the provided nomenclature rule and paper-local evidence.

Evaluate all seven slots, including slots whose module and semantic_value are null:
- lineage
- function
- migration
- migration_subscript
- differentiation
- differentiation_subscript
- antigen

A null slot is correct when neither the original label nor the supplied evidence supports a value for that slot. Do not use insufficient_evidence merely because a slot is null; use it only when the supplied evidence is genuinely ambiguous or inadequate to verify a biologically relevant claim.

For each slot, return exactly these fields:
- verdict: "yes" or "no"
- error_type: null when verdict is "yes"; otherwise exactly one allowed error type
- reason: null when verdict is "yes"; otherwise one concise explanation grounded in the rule and evidence

Allowed error types:
- unsupported_value: the mapped slot adds a biological property not supported by the original label or evidence
- missing_value: the evidence supports a property that the mapped slot omits
- incorrect_normalization: the biological meaning is supported, but module or semantic_value does not follow the rule
- wrong_slot: supported information is assigned to the wrong slot
- insufficient_evidence: the available paper-local evidence is insufficient to verify the slot

When information is assigned to the wrong slot, mark that populated slot as wrong_slot and mark the slot that should contain the information as missing_value. This ensures both affected slots are identifiable.

Do not generate revised modules, revised semantic values, a corrected mapped label, or any correction instructions. Judge only the supplied mapping.

Return only one valid JSON object with the exact top-level field "slot_verdicts". Return every slot exactly once and do not add fields.

Output schema:
{
  "slot_verdicts": {
    "lineage": {"verdict": "yes", "error_type": null, "reason": null},
    "function": {"verdict": "yes", "error_type": null, "reason": null},
    "migration": {"verdict": "yes", "error_type": null, "reason": null},
    "migration_subscript": {"verdict": "yes", "error_type": null, "reason": null},
    "differentiation": {"verdict": "yes", "error_type": null, "reason": null},
    "differentiation_subscript": {"verdict": "yes", "error_type": null, "reason": null},
    "antigen": {"verdict": "yes", "error_type": null, "reason": null}
  }
}

NOMENCLATURE RULE:
<<<
<T-CELL NOMENCLATURE REFERENCE>
>>>

MAPPING RECORD:
<MAPPING RECORD JSON>
\end{promptlisting}

\section{Paper Extracted Label to Data Annotation Prompt}
\label{sec:supp-paper-data-matching}

\begin{promptlisting}{Annotation Matching}

You are matching PDF-extracted cell type names to data annotation labels from the same paper.

MATCHING RULES, in priority order:
1. CLUSTER ID (authoritative). If the extracted label and a candidate both contain a cluster identifier (e.g. "c0", "c11", "cluster 3"), match on that ID alone. It overrides all semantic similarity. Never match differing IDs.
2. LINEAGE CONSTRAINT (hard block). If the extracted label specifies a lineage(CD4, CD8, Treg, TFH, NKT, MAIT, gamma-delta) and no candidate shares it, return NO_MATCH. Never map across a specified lineage.
3. MARKER EVIDENCE. Candidate labels often embed markers (CCR7, FOXP3, GZMK, MKI67). Match these against the extracted label's text and evidence.
4. SEMANTIC CORRESPONDENCE. Prefer specific biological correspondence over loose family-level similarity.

Choose EXACTLY ONE candidate, or NO_MATCH. Copy the chosen candidate label EXACTLY
as written.

NO_MATCH is a valid, expected answer. Return NO_MATCH when:
 - no candidate plausibly refers to the same population, or
 - the only candidates conflict on lineage (rule 2), or
 - the population was likely filtered out of this data object.
Do NOT force a match. A wrong match is worse than NO_MATCH.

WHEN THE EXTRACTED LABEL SPANS SEVERAL CANDIDATES (e.g. "proliferating T cells" vs four cycling subclusters), still choose ONE: the closest biological representative, preferring (a) shared cluster ID, then (b) the candidate whose markers appear in the evidence, then (c) the largest/most canonical member of that program. Set "spans_multiple": true and name the others in "reason".

Ground every decision in the extracted label text and its evidence.

Return JSON only:
{"items":[{"original_label": "...",
           "matched_data_label": "..." | null,
           "match_status": "cluster_id" | "marker" | "semantic" | "no_match",
           "spans_multiple": true | false,
           "confidence": "high" | "medium" | "low",
           "reason": "short, concrete"}]}


\end{promptlisting}

\section{Technical Details of Compared Baselines}
\label{apex:baselines}

We compare \model with existing cross-study cell population alignment approaches.


\noindent \textbf{BM25.} BM25 (Best Matching 25) ~\cite{robertson2009BM25} is a ranking function used in information retrieval to determine how relevant a document is to a given query. 
We cast cell type normalization as retrieval over the Cell Ontology. We treat each paper-local label as a query and each CL surface form as a document. Every label is scored against all CL surface forms and the top-ranked term is returned as its normalized label.

\noindent \textbf{Levenshtein.} Levenshtein distance ~\cite{ristad1998editDist} measures the difference between two strings as the minimum number of character edits required to transform one string into the other. We compute the distance between each extracted paper-local labels and every CL surface form, and convert it into a length-normalized similarity to ensure candidate names of different lengths remain comparable. The CL term with highest similarity is returned as its normalized label.  

\noindent \textbf{PubMedBERT.} PubMedBERT ~\cite{gu2022pubmedbert} is a domain-specific language model developed by Microsoft Research pretrained on biomedical text from PubMed abstracts and full-text articles. We use it as a frozen encoder without task-specific fine-tuning. Each extracted paper-local label and each CL label is encoded into a fixed-length embedding, and every label is assigned to the CL term with the highest cosine similarity in the shared embedding space.

\noindent \textbf{SapBERT.} SapBERT ~\cite{liu2021SapBERT} is a specialized language model that links different names of the same medical concept by aligning their vector representations in the embedding space.  It is initialized from PubMedBERT and further pretrained on synonym pairs drawn from the UMLS. Similar to PubMedBERT, we use SapBERT to encode paper-local labels and CL surface forms and assign each label to the CL term with the highest cosine similarity. The two baselines therefore differ only in this additional self-alignment pretraining, and their comparison isolates its effect on cell type normalization.

\noindent \textbf{ZOOMA.}
ZOOMA~\cite{zooma} is an ontology annotation service developed and maintained by the European Bioinformatics Institute (EMBL-EBI). It integrates dictionary-based lookup, curated annotation resources, ontology synonyms, and semantic matching to retrieve candidate ontology terms for a given biological entity. We apply ZOOMA to the extracted paper-local labels and configure the target ontologies to include Cell Ontology, Cell Line Ontology, Human Cell Atlas Ontology, and Provisional Cell Ontology.

\noindent \textbf{BioPortal.}
BioPortal~\cite{bioportal} provides ontology-aware search across a large collection of biomedical ontologies. Given an extracted paper-local label, BioPortal retrieves candidate ontology concepts based on lexical similarity and ontology metadata. We query BioPortal against multiple cell-related ontologies, including Cell Ontology, Cell Line Ontology, Placental Cell Type Ontology, Brain Region \& Cell Type Terminology, and Breast Tissue Cell Lines Ontology.


\section{Technical Details of Evaluation Metrics}
\label{apex:evaluation-metrics}

\subsection{Annotation-based Evaluation}

Given the predicted normalization mapping $f_{\mathrm{pred}}:O\rightarrow M$, where $M=\{m_1,m_2,\ldots,m_n\}$ denotes the normalized labels generated by a mapping method, and the validated reference mapping $f_{\mathrm{ref}}:O\rightarrow CL$ obtained from the standardized Cell Ontology annotations in CZ CELLxGENE, we propagate both mappings through the shared paper-local labels. Consequently, each paper-local label is associated with both a reference Cell Ontology label $CL$ and a predicted normalized label $M$.
Rather than evaluating exact string matching, we assess whether the predicted normalization induces a partition that is consistent with the reference Cell Ontology annotations. 

\noindent \textbf{Adjusted Mutual Information (AMI).}
We first compute the Adjusted Mutual Information (AMI)~\cite{vinh2010information} between the predicted labels $M$ and the reference Cell Ontology labels $CL$. Let $I(CL,M)$ denote the mutual information between the two partitions. AMI is defined as
\begin{equation}
\mathrm{AMI}(CL,M)
=
\frac{
I(CL,M)-\mathbb{E}[I(CL,M)]
}{
\max\{H(CL),H(M)\}-\mathbb{E}[I(CL,M)]
},
\end{equation}
where $H(\cdot)$ denotes entropy and $\mathbb{E}[I(CL,M)]$ is the expected mutual information under random assignments. AMI can take negative values when agreement is below chance, with larger values indicating stronger agreement after correcting for chance.

\noindent \textbf{Merge F1.} While AMI evaluates the overall agreement between two partitions, it does not directly assess whether a normalization method makes biologically correct merge decisions. We therefore propose to evaluate the correctness of merge operations by considering every pair of paper-local labels.
For every pair $(o_i,o_j)$, we define
\begin{equation}
\delta_M(o_i,o_j)=
\begin{cases}
1,&
f_{\mathrm{pred}}(o_i)=f_{\mathrm{pred}}(o_j),\\
0,&
\text{otherwise},
\end{cases}
\end{equation}
, and similarly
\begin{equation}
\delta_{CL}(o_i,o_j)=
\begin{cases}
1,&
f_{\mathrm{ref}}(o_i)=f_{\mathrm{ref}}(o_j),\\
0,&
\text{otherwise}.
\end{cases}
\end{equation}
Based on these pairwise relationships, we define
\begin{itemize}
    \item True Positive (TP): $\delta_M=1$ and $\delta_{CL}=1$;
    \item False Positive (FP): $\delta_M=1$ but $\delta_{CL}=0$;
    \item False Negative (FN): $\delta_M=0$ but $\delta_{CL}=1$.
\end{itemize}
Precision, Recall, and Merge F1 are computed as
\begin{subequations}
\begin{align}
\mathrm{Precision}&= \frac{TP}{TP+FP}, \\
\mathrm{Recall}&=\frac{TP}{TP+FN}, \\
\mathrm{Merge\ F1}&= \frac{
2\cdot\mathrm{Precision}\cdot\mathrm{Recall}
}{
\mathrm{Precision}+\mathrm{Recall}}.
\end{align}
\end{subequations}
Unlike AMI, Merge F1 directly evaluates whether a normalization method correctly merges biologically equivalent paper-local labels while avoiding incorrect merges across distinct Cell Ontology classes.

\subsection{Expression-based Evaluation.}

Although the annotation-based metrics quantify semantic agreement with Cell Ontology, they do not evaluate whether the merged labels correspond to transcriptionally coherent cellular populations. We therefore perform a secondary evaluation using gene expression embeddings.
For each predicted normalized label $m$, suppose it contains $k$ author-provided annotations,
\begin{equation}
    A(m)=\{a_1,a_2,\ldots,a_k\}.
\end{equation}
For each annotation $a_i$, we compute its centroid in the embedding space,
\begin{equation}
c_i =
\frac{1}{|a_i|}
\sum_{x\in a_i}
x.
\end{equation}
The biological consistency of the merged label is quantified by the average pairwise cosine similarity among these annotation centroids,
\begin{equation}
\mathrm{MC}(m)
=
\frac{2}{k(k-1)}
\sum_{i<j}
\cos(c_i,c_j),
\end{equation}
where $\cos(\cdot,\cdot)$ denotes cosine similarity.
Finally, the overall Merge Consistency score is computed as
\begin{equation}
\mathrm{MergeConsistency}
=
\frac{1}{|\mathcal{M}|}
\sum_{m\in\mathcal{M}}
\mathrm{MC}(m),
\end{equation}
where $\mathcal{M}$ denotes the set of normalized labels that merge two or more author-provided annotations. Higher Merge Consistency indicates that the author-provided annotations merged by a normalization method exhibit high transcriptomic similarity, providing biological evidence that the merge is plausible.

\subsection{Soft Bipartite Matching F1 Score}\label{appex:SBMS-F1}
SBMS-F1 proceeds as follows. Let the gold-standard cell type set be $O={o_1, o_2, \dots, o_n}$, and the human or model extracted cell type set be $P={p_1, p_2, \dots, p_m}$. Each cell type surface text is first embedded using Qwen3-Embedding-8B. A pairwise cosine similarity matrix $\mathbf{S}\in [0,1]^{n\times m}$ is constructed with entries
\begin{equation}
    s_{ij}=\cos (\mathbf{e}^{O}_{i},\mathbf{e}^{P}_{j}),
\end{equation}
where $\mathbf{e}^{O}_{i}$ and $\mathbf{e}^{P}_{j}$ are the embedding vectors of the i-th gold term and the j-th predicted term, respectively.
Next, the Hungarian algorithm~\cite{kuhn1955hungarian} is applied to find the optimal one‑to‑one assignment that maximizes the total similarity, \textit{i.e.},
\begin{equation}
    \pi^{*}=\arg \max_{\pi} \sum_i^{n}s_{i,\pi(i)},
\end{equation}
where $\pi$ is a matching that pairs each gold term with at most one predicted term. This transforms the evaluation into a bipartite graph matching problem.
Finally, the summed similarity of the matched pairs is normalized to obtain precision, recall, and the harmonic mean F1 score:
\begin{subequations}
\begin{align}
\text{SBMS-P} &= \frac{1}{m}\sum_i^{n} s_{i,\pi^{*}(i)}, \\
\text{SBMS-R} &= \frac{1}{n}\sum_i^{n} s_{i,\pi^{*}(i)}, \\
\text{SBMS-F1} &= \frac{2\cdot\text{SBMS-P}\cdot\text{SBMS-R}}{\text{SBMS-P}+\text{SBMS-R}},
\end{align}
\end{subequations}
with the convention $\text{SBMS-P}=0$ if $m=0$, and $\text{SBMS-R}=0$ if $n=0$.
This design allows SBMS-F1 to reward semantically correct but lexically different outputs (e.g., ``natural killer T cells'' vs.\ ``NKT-like cells'') while penalizing missing or spurious extractions. The metric is inherently soft, bounded between $0$ and $1$, and reduces to the conventional exact‑match $F_1$ when all matched pairs are identical (i.e., when every $s_{ij}=1$ for the assigned pairs).

\begin{table*}[t]
  \caption{Per-study data scale and T-cell label counts across the final 44-paper analysis set.}
  \label{tab:supp-per-study-summary}
  \centering
  \scriptsize
  \setlength{\tabcolsep}{2.8pt}
  \renewcommand{\arraystretch}{1.05}
  \begin{tabularx}{\textwidth}{@{}r >{\raggedright\arraybackslash}X l r r rr rr@{}}
  \toprule
  & & & & & \multicolumn{2}{c}{\model-GPT} & \multicolumn{2}{c}{\model-DS} \\
  \cmidrule(lr){6-7}\cmidrule(lr){8-9}
  No. & Paper title & Category & \shortstack{Cells\\($n$)} & \shortstack{Annotated cell\\types ($n$)} & Extracted & Mapped & Extracted & Mapped \\
  \midrule
  1 & A single-cell and spatially resolved atlas of human breast cancers. & Cancer & 35,214 & 12 & 10 & 8 & 10 & 8 \\
  2 & Spatially organized multicellular immune hubs in human colorectal cancer. & Cancer & 76,965 & 26 & 9 & 6 & 7 & 3 \\
  3 & Single-cell transcriptomics of human T cells reveals tissue and activation signatures in health and disease. & Cancer & 51,876 & 6 & 18 & 14 & 10 & 10 \\
  4 & Characteristics of anti-CD19 CAR T cell infusion products associated with efficacy and toxicity in patients with large B cell lymphomas. & Cancer & 133,405 & 15 & 6 & 5 & 3 & 3 \\
  5 & High-resolution single-cell atlas reveals diversity and plasticity of tissue-resident neutrophils in non-small cell lung cancer. & Cancer & 230,697 & 9 & 8 & 7 & 8 & 7 \\
  6 & Signatures of plasticity, metastasis, and immunosuppression in an atlas of human small cell lung cancer. & Cancer & 46,140 & 6 & 6 & 6 & 6 & 6 \\
  7 & Stromal cell diversity associated with immune evasion in human triple-negative breast cancer. & Cancer & 7,990 & 7 & 8 & 7 & 6 & 6 \\
  8 & Single-cell sequencing links multiregional immune landscapes and tissue-resident T cells in ccRCC to tumor topology and therapy efficacy. & Cancer & 73,806 & 10 & 10 & 8 & 10 & 10 \\
  9 & Ovarian cancer mutational processes drive site-specific immune evasion. & Cancer & 221,315 & 41 & 18 & 13 & 18 & 12 \\
  10 & Mapping single-cell transcriptomes in the intra-tumoral and associated territories of kidney cancer. & Cancer & 141,106 & 26 & 16 & 12 & 9 & 7 \\
  11 & Single-cell proteo-genomic reference maps of the hematopoietic system enable the purification and massive profiling of precisely defined cell states. & Cancer & 4,048 & 9 & 11 & 10 & 10 & 5 \\
  12 & A single-cell atlas enables mapping of homeostatic cellular shifts in the adult human breast. & Cancer & 17,483 & 6 & 9 & 7 & 6 & 5 \\
  13 & Follicular Lymphoma Microenvironment Characteristics Associated with Tumor Cell Mutations and MHC Class II Expression. & Cancer & 35,483 & 11 & 11 & 11 & 7 & 7 \\
  14 & Single cell RNA sequencing of human liver reveals distinct intrahepatic macrophage populations & Healthy & 1,786 & 3 & 3 & 2 & 3 & 3 \\
  15 & Cross-tissue immune cell analysis reveals tissue-specific features in humans & Healthy & 216,611 & 18 & 16 & 15 & 12 & 11 \\
  16 & Distinct microbial and immune niches of the human colon & Healthy & 18,574 & 9 & 14 & 11 & 13 & 10 \\
  17 & A spatially resolved single-cell genomic atlas of the adult human breast & Healthy & 76,567 & 14 & 12 & 12 & 10 & 10 \\
  18 & Human skeletal muscle aging atlas & Healthy & 6,875 & 6 & 4 & 3 & 4 & 3 \\
  19 & Blood and immune development in human fetal bone marrow and Down syndrome & Healthy & 1,349 & 8 & 4 & 4 & 4 & 4 \\
  20 & Asian diversity in human immune cells. & Healthy & 630,055 & 25 & 11 & 10 & 8 & 8 \\
  21 & A spatial human thymus cell atlas mapped to a continuous tissue axis & Healthy & 357,049 & 26 & 16 & 8 & 14 & 5 \\
  22 & Multi-omic profiling reveals age-related immune dynamics in healthy adults & Healthy & 997,874 & 35 & 48 & 27 & 43 & 22 \\
  23 & Single-cell and spatial mapping identify cell types and signaling networks in the human ureter & Healthy & 6,045 & 8 & 8 & 6 & 7 & 6 \\
  24 & Multimodal profiling reveals tissue-directed signatures of human immune cells altered with age & Healthy & 610,429 & 48 & 15 & 11 & 13 & 11 \\
  25 & Cellular heterogeneity and dynamics of the human uterus in healthy premenopausal women & Healthy & 3,503 & 3 & 4 & 4 & 4 & 4 \\
  26 & A single-cell atlas of the peripheral immune response in patients with severe COVID-19. & Infectious & 15,607 & 6 & 4 & 4 & 3 & 3 \\
  27 & Immunophenotyping of COVID-19 and influenza highlights the role of type I interferons in development of severe COVID-19. & Infectious & 15,790 & 4 & 7 & 5 & 6 & 6 \\
  28 & COVID-19 immune features revealed by a large-scale single-cell transcriptome atlas. & Infectious & 634,595 & 28 & 15 & 8 & 13 & 8 \\
  29 & Single-cell multi-omics analysis of the immune response in COVID-19. & Infectious & 286,085 & 17 & 17 & 17 & 17 & 14 \\
  30 & Local and systemic responses to SARS-CoV-2 infection in children and adults. & Infectious & 10,385 & 16 & 21 & 17 & 23 & 15 \\
  31 & A blood atlas of COVID-19 defines hallmarks of disease severity and specificity. & Infectious & 402,103 & 21 & 28 & 17 & 20 & 13 \\
  32 & Time-resolved systems immunology reveals a late juncture linked to fatal COVID-19. & Infectious & 204,133 & 11 & 14 & 12 & 13 & 9 \\
  33 & Type I interferon autoantibodies are associated with systemic immune alterations in patients with COVID-19. & Infectious & 275,099 & 14 & 6 & 6 & 3 & 3 \\
  34 & Impaired local intrinsic immunity to SARS-CoV-2 infection in severe COVID-19. & Infectious & 1,475 & 3 & 3 & 3 & 3 & 3 \\
  35 & Early human lung immune cell development and its role in epithelial cell fate. & Infectious & 11,123 & 6 & 6 & 6 & 6 & 6 \\
  36 & Interstitial macrophages are a focus of viral takeover and inflammation in COVID-19 initiation in human lung. & Infectious & 18,272 & 10 & 9 & 9 & 8 & 8 \\
  37 & An interactive single cell web portal identifies gene and cell networks in COVID-19 host responses. & Infectious & 109,995 & 15 & 11 & 11 & 11 & 11 \\
  38 & mRNA COVID-19 vaccine elicits potent adaptive immune response without the acute inflammation of SARS-CoV-2 infection. & Infectious & 127,655 & 13 & 5 & 5 & 5 & 5 \\
  39 & Intra- and Inter-cellular Rewiring of the Human Colon during Ulcerative Colitis. & Inflammatory & 1,293 & 9 & 16 & 9 & 11 & 7 \\
  40 & Cells of the human intestinal tract mapped across space and time. & Inflammatory & 34,764 & 18 & 18 & 13 & 17 & 12 \\
  41 & Single-cell RNA-seq reveals cell type-specific molecular and genetic associations to lupus. & Inflammatory & 551,388 & 7 & 7 & 7 & 7 & 7 \\
  42 & The landscape of immune dysregulation in Crohn's disease revealed through single-cell transcriptomic profiling in the ileum and colon. & Inflammatory & 78,818 & 8 & 6 & 5 & 5 & 5 \\
  43 & Single-cell integration reveals metaplasia in inflammatory gut diseases. & Inflammatory & 262,642 & 17 & 11 & 10 & 10 & 9 \\
  44 & Single-cell and spatially resolved interactomics of tooth-associated keratinocytes in periodontitis. & Inflammatory & 19,840 & 7 & 10 & 10 & 10 & 10 \\
  \bottomrule
  \end{tabularx}

  \parbox{\textwidth}{\scriptsize\textit{Notes.} Cells are observations in the selected T-cell-focused H5AD dataset for each study. ``Annotated cell types'' denotes the number of distinct non-missing values in the author-provided annotation field used for analysis. Extracted counts are distinct Stage~1 paper-local labels after exact duplicate removal. Mapped counts are distinct non-empty Stage~2 normalized labels; several extracted labels may map to one normalized label. GPT and DS denote the final GPT-5.5 and DeepSeek-V4-Pro (thinking) configurations, respectively.}
\end{table*}

\begin{table*}[t]
\centering
\caption{Example mappings between paper-reported labels, dataset annotations, normalized labels, and Cell Ontology terms.}
\label{tab:case-study}

\small
\begin{tabularx}{\textwidth}{>{\raggedright\arraybackslash}X
>{\raggedright\arraybackslash}X
c
>{\raggedright\arraybackslash}X
c}
\toprule
\textbf{Paper Label} &
\textbf{Dataset Annotation} &
\textbf{Mapped Label} &
\textbf{Cell Ontology Label} &
\textbf{CL ID} \\
\midrule

naïve/central memory CD4$^{+}$ T cells (CCR7/c0)
&
T\_cells\_c0\_CD4+\_CCR7
&
CD4+ TS
&
CD4-positive, alpha-beta T cell
&
CL:0000624
\\

Th1 CD4 effector memory T cells (IL7R/c1)
&
T\_cells\_c1\_CD4+\_IL7R
&
CD4+ TH1DM
&
CD4-positive, alpha-beta T cell
&
CL:0000624
\\

FOXP3$^{+}$ regulatory T cells (FOXP3/c2)
&
T\_cells\_c2\_CD4+\_T-regs\_FOXP3
&
CD4 Treg
&
CD4-positive, alpha-beta T cell
&
CL:0000624
\\

T follicular helper cells (CXCL13/c3)
&
T\_cells\_c3\_CD4+\_Tfh\_CXCL13
&
CD4 TFH
&
CD4-positive, alpha-beta T cell
&
CL:0000624
\\

Chemokine-expressing CD8$^{+}$ T cells (ZFP36/c4)
&
T\_cells\_c4\_CD8+\_ZFP36
&
CD8+ T
&
CD8-positive, alpha-beta T cell
&
CL:0000625
\\

Type I interferon signature T cells (IFIT1/c6)
&
T\_cells\_c6\_IFIT1
&
T
&
CD8-positive, alpha-beta T cell
&
CL:0000625
\\

PDCD1\textsuperscript{low} CD8$^{+}$ T cells expressing IFNG and TNF (IFNG/c7)
&
T\_cells\_c7\_CD8+\_IFNG
&
CD8+ T
&
CD8-positive, alpha-beta T cell
&
CL:0000625
\\

Exhausted CD8 T cells (LAG3/c8)
&
T\_cells\_c8\_CD8+\_LAG3
&
CD8+ TX
&
CD8-positive, alpha-beta T cell
&
CL:0000625
\\

NKT-like cells (FCGR3A/c10)
&
T\_cells\_c10\_NKT\_cells\_FCGR3A
&
NKT
&
Mature NK T cell
&
CL:0000814
\\

Proliferating T cells (MKI67/c11)
&
T\_cells\_c11\_MKI67
&
T
&
T cell
&
CL:0000084
\\

\bottomrule
\end{tabularx}
\end{table*}

\begin{table*}[t]
  \caption{Seven-paper evaluation set for the Stage~1 Information Extraction Agent.}
  \label{tab:supp-stage1-paper-inventory}
  \centering
  \scriptsize
  \setlength{\tabcolsep}{3.2pt}
  \renewcommand{\arraystretch}{1.14}
  \begin{tabularx}{\textwidth}{@{}r X l c l c@{}}
  \toprule
  No. & Paper title & Venue & Year & DOI & \shortstack{In 44-study\\benchmark} \\
  \midrule
  1 & A single-cell and spatially resolved atlas of human breast cancers~\cite{wu2021breastatlas} & \textit{Nature Genetics} & 2021 & \url{https://doi.org/10.1038/s41588-021-00911-1} & Yes \\
  2 & Integrative single-cell multi-omics of CD19-CAR\textsuperscript{pos} and CAR\textsuperscript{neg} T cells suggest drivers of immunotherapy response in B-cell neoplasias~\cite{guerreromurillo2024cartmultiomics} & \textit{bioRxiv} & 2024 & \url{https://doi.org/10.1101/2024.01.23.576878} & No \\
  3 & Pan-cancer T cell atlas links a cellular stress response state to immunotherapy resistance~\cite{chu2023pancancertcellatlas} & \textit{Nature Medicine} & 2023 & \url{https://doi.org/10.1038/s41591-023-02371-y} & No \\
  4 & Follicular Lymphoma Microenvironment Characteristics Associated with Tumor Cell Mutations and MHC Class II Expression~\cite{han2022follicularlymphoma} & \textit{Blood Cancer Discovery} & 2022 & \url{https://doi.org/10.1158/2643-3230.BCD-21-0075} & Yes \\
  5 & Single-cell transcriptomic analysis of gingivo-buccal oral cancer reveals two dominant cellular programs~\cite{kurkalang2023oralcancer} & \textit{Cancer Science} & 2023 & \url{https://doi.org/10.1111/cas.15979} & No \\
  6 & Ovarian cancer mutational processes drive site-specific immune evasion~\cite{vazquezgarcia2022ovarian} & \textit{Nature} & 2022 & \url{https://doi.org/10.1038/s41586-022-05496-1} & Yes \\
  7 & Stromal cell diversity associated with immune evasion in human triple-negative breast cancer~\cite{wu2020tnbcstroma} & \textit{The EMBO Journal} & 2020 & \url{https://doi.org/10.15252/embj.2019104063} & Yes \\
  \bottomrule
  \end{tabularx}
\end{table*}

\end{document}